\documentclass[superscriptaddress,10pt,aps,prl,twocolumn,longbibliography,notitlepage]{revtex4-1}
\usepackage{float}
\usepackage{array}
\usepackage{graphicx}
\usepackage{amsbsy}
\usepackage{epstopdf}
\usepackage{amsmath,amssymb,amsfonts,amsthm}
\usepackage{indentfirst}
\usepackage[T1]{fontenc}
\usepackage[dvipsnames]{xcolor}
\usepackage{url}
\usepackage[colorlinks]{hyperref}
\usepackage{etoolbox}
\hypersetup{
    plainpages=true,
    breaklinks=true,
    hypertexnames=false,
    pageanchor=true,
    colorlinks=true,
    linkcolor={blue},
    citecolor={red},
    urlcolor={blue},
    anchorcolor={black}
}

\makeatletter
\newcommand{\startsupplementtoc}{%
  \newwrite\@supp@tocfile
  \immediate\openout\@supp@tocfile=\jobname.suptoc
  \let\oldaddcontentsline\addcontentsline
  \renewcommand{\addcontentsline}[3]{%
    \immediate\write\@supp@tocfile{%
      \string\contentsline{##1}{##2}{\thepage}{}%
    }%
  }%
}

\newcommand{\printsupplementtoc}{%
  \begingroup
  \let\contentsline\oldcontentsline
  \input{\jobname.suptoc}
  \endgroup
}
\makeatother

\newcommand{\suppressTOC}{%
  \let\oldaddcontentsline\addcontentsline
  \renewcommand{\addcontentsline}[3]{}%
}

\newcommand{\restoreTOC}{%
  \let\addcontentsline\oldaddcontentsline
}

\definecolor{myred}{rgb}{1,0,0}

\renewcommand{\figurename}{\textbf{Fig.}}
\renewcommand{\thefigure}{\textbf{\arabic{figure}}}
\begin{document}

\title{Observation of Restored Adiabatic State Transfer in Time-Modulated Non-Hermitian Systems}

\author{Xiaowei Wang}\thanks{These authors contributed equally to this work.}
\affiliation{Beijing Computational Science Research Center, Beijing 100193, China}
\author{Ievgen I. Arkhipov}\thanks{These authors contributed equally to this work.}
\affiliation{Joint Laboratory of Optics of Palack\'y University and Institute of Physics of CAS, Faculty of Science, Palack\'y University, 17. listopadu 12, 771 46 Olomouc, Czech Republic}
\author{Quan Lin}
\author{Huixia Gao}
\author{Dengke Qu}
\author{Lei Xiao}
\affiliation{School of Physics, Southeast University, Nanjing 211189, China}
\author{Franco Nori}
\affiliation{Quantum Information Physics Theory Research Team, Quantum Computing Center, RIKEN, Wakoshi, Saitama, 351-0198, Japan}
\author{Peng Xue}\email{gnep.eux@gmail.com}
\affiliation{School of Physics, Southeast University, Nanjing 211189, China}

\begin{abstract}
{\bf Exceptional points (EPs) have attracted extensive research interest due to their intriguing properties. One of the  hallmarks of EP physics is that dynamically encircling the EPs induces chiral mode switching, arising from the breakdown of adiabaticity due to the presence of a complex spectrum in the system's Hamiltonian. While such chiral mode behavior has been widely observed experimentally, achieving truly adiabatic, and thus symmetric, state transfer, regardless of the winding direction, in time-modulated non-Hermitian systems has remained elusive.
In this work, we demonstrate that this long-sought
adiabatic state dynamics can indeed be restored. By steering a two-mode photonic setup along specifically designed trajectories in parameter space, we realize conditions where the associated non-Hermitian evolution operator acquires a purely real spectrum.
Moreover, our experimental platform enables controlled switching between symmetric (adiabatic) and chiral (non-adiabatic) state-transfer regimes for the same set of initial modes, thus effectively implementing a universal symmetric-asymmetric two-mode switch. Our results therefore open new avenues for harnessing unique topological spectral properties of non-Hermitian systems,
paving the way for the practical design of versatile optical wave-manipulation devices and for advancing both classical and quantum information technologies.}
\end{abstract}

\date{\today}

\maketitle

\suppressTOC

Non-Hermitian systems with gain and loss exhibit rich physical phenomena that are fundamentally distinct from Hermitian quantum systems, attracting broad interest in current research~\cite{RKM+18}. Their inherently complex energy spectra give rise to unique spectral degeneracies known as EPs~\cite{MR08,MA19,IASK18}, which are characterized by the simultaneous coalescence of both eigenvalues and their corresponding eigenvectors. The topological structure near EPs underlies a variety of counterintuitive phenomena~\cite{lin2011unidirectional,FXFF+13,PZZ+16,liu2016metrology,HMH+14,BSS+14,schumer2022topological,HUG+17,chen2017exceptional,zhang2019quantum,Wu2025,song2024experimental}, ranging from nonreciprocal optoelectronic devices~\cite{lin2011unidirectional,FXFF+13,PZZ+16} and unconventional lasing effects~\cite{HMH+14,BSS+14,schumer2022topological} to enhanced sensing~\cite{HUG+17,chen2017exceptional,zhang2019quantum}, to name a few (see also reviews~\cite{Ozdemir2019,Ashida2020}).

Exceptional points endow non-Hermitian spectra with a characteristic topological structure that can profoundly affect their state evolution. In particular, when system parameters trace a closed loop around an EP, the associated Riemann-sheet topology can induce a symmetric permutation of eigenstates, such that an eigenmode does not return to itself after a single cycle and the final state depends solely on the initial condition, irrespective of the winding direction. This symmetric state-exchange effect has been predicted theoretically~\cite{Heiss2000,Cartarius2007} and observed experimentally~\cite{Dembowski2001,Dietz2011,Gao2015,Ding2016,Ergoktas2022,Guria2024} using quasi-static, namely stroboscopic, protocols, where eigenstates are independently probed at successive points in parameter space without invoking continuous time evolution.

When the same EP-encircling is implemented dynamically, however, qualitatively different behavior emerges. A prominent consequence is chiral mode transfer:  when dynamically encircling an EP, the final state depends solely on the encircling direction regardless of the initial state~\cite{Doppler2016,Hassan2017,Xu2016,YZZ+18,Zhang2018_encirc,Zhang2019_encirc2,Feilhauer2020,LWD+21,Feng2022,Tang2023,GSQ+25}.
The complex Riemann surface structure in the system's spectrum underlies this chiral behavior, namely, the eigenstate with the larger imaginary part of the eigenvalue is preferentially populated due to non-adiabatic transitions (NATs) which occur during the state evolution~\cite{Doppler2016}. While this chiral mode behavior has been extensively studied both theoretically~\cite{Uzdin2011,Berry2011,Graefe2013} and experimentally~\cite{Doppler2016,Hassan2017,Xu2016,YZZ+18,Zhang2018_encirc,Zhang2019_encirc2,Feng2022,Chen2022,Tang2023}, achieving truly adiabatic and symmetric state transfer in time-modulated non-Hermitian systems has remained a formidable challenge.

Adiabatic and symmetric state transfer can be restored in non-Hermitian systems, as demonstrated by theoretical study~\cite{AMI+24}. {In this context, the restored adiabaticity refers to the absence of discontinuous NATs during the state evolution.} Such behavior is realized by dynamically encircling an EP along suitably designed trajectories in parameter space, for which the non-Hermitian evolution operator exhibits a purely real spectrum. The resulting suppression of NATs thus enables smooth and symmetric state transfer. This behavior is in sharp contrast to previously studied dynamical EP-encircling protocols, where complex spectra generically trigger non-adiabatic jumps and give rise to chiral, direction-dependent state transfer~\cite{Doppler2016,Hassan2017}.

\begin{figure*}
    \includegraphics[width=0.95\textwidth]{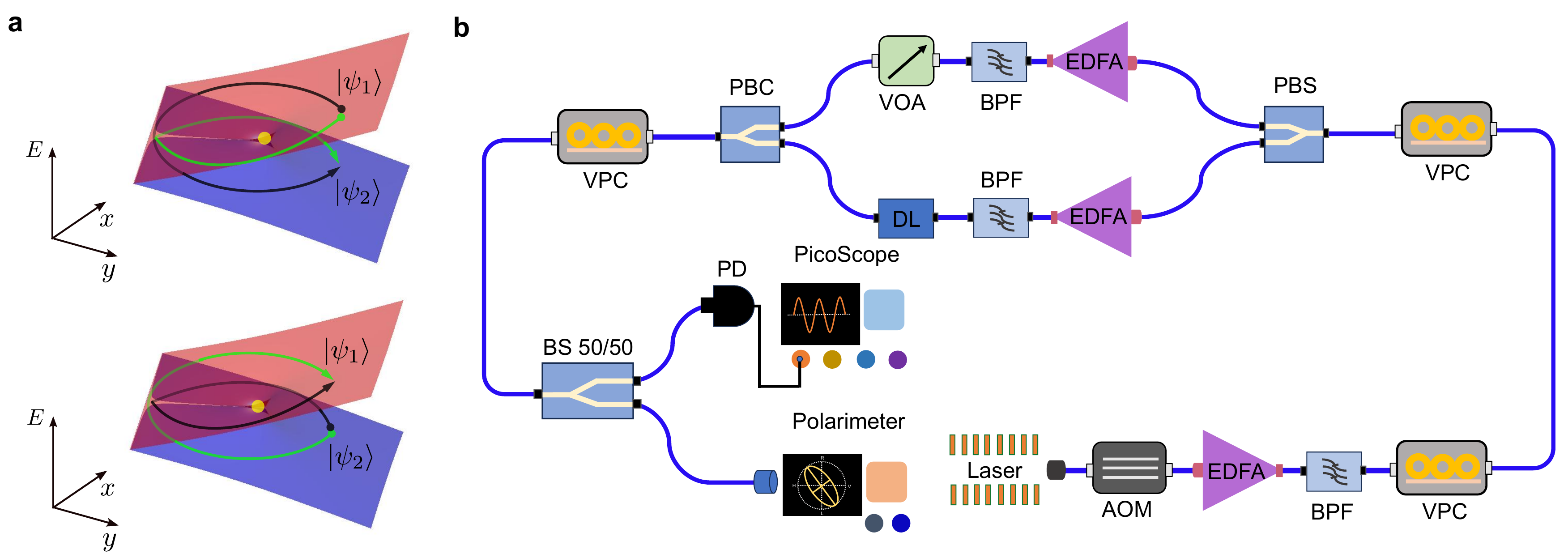}
    \caption{{\bf Schematic illustration of the experimental implementation.} {\bf a} Schematic diagram of adiabatic evolution by dynamically encircling the EP in a two-level non-Hermitian system. We present schematic trajectories with different encircling directions starting from the two involved Riemann sheets, where the red and blue surfaces represent the two eigenenergies. The yellow sphere represents the EP point. The state transfer is observed to be independent of the direction of the encirclement, where the green/black trajectories denote clockwise/counterclockwise encirclement.
    {\bf b} Experimental setup. The seed pulse module consists of a distributed feedback laser and an acousto-optic modulator (AOM) for generating pulses, an erbium-doped fiber amplifier (EDFA) for pulse amplification, a narrow bandpass filter (BPF) to suppress the excessive noise introduced by the EDFA, and a variable  polarization controller (VPC) for preparing the initial polarization state. Two VPCs are then used to couple the polarization components to implement unitary operations. These non-unitary operations are realized by using the polarized beam splitter (PBS), EDFAs, the optical delay line (DL), a variable optical attenuator (VOA), and the polarized beam combiner (PBC), where the horizontal and vertical polarization components can be independently controlled along two separate optical paths. Finally, half of the pulses are routed to a monitoring module equipped with an InGaAs photodetector (PD) and a PicoScope, and the remaining pulses are used for polarization state analysis via a polarimeter.}
    \label{figure1}
\end{figure*}

Here we report the experimental observation of  adiabatic and symmetric state transfer in a time-modulated non-Hermitian photonic platform. By monitoring the polarization state of a laser pulse under controlled intensity, phase, and polarization modulations, we realize specific closed loops in the parameter space that dynamically encircle an EP. Along these specially chosen trajectories, we demonstrate the full restoration of adiabaticity, independent of the direction of encirclement.

Crucially, the same platform also enables us to switch between symmetric (adiabatic) and chiral (non-adiabatic) state-transfer regimes for a fixed initial set of eigenstates. This constitutes the effective realization of a programmable symmetric–chiral state switch, a capability that was previously considered unattainable in non-Hermitian two-level systems. Specifically, for a given initial and final Hamiltonian, the nature of the state transfer (symmetric or asymmetric) can be selected solely through the proper choice of parameter modulation within a single physical platform.
This work provides an alternative route for the dynamical control of states in non-Hermitian systems, providing opportunities for next-generation optical wave-manipulation devices and advancing both classical and quantum technologies.

\begin{figure*}
    \includegraphics[width=0.99\textwidth]{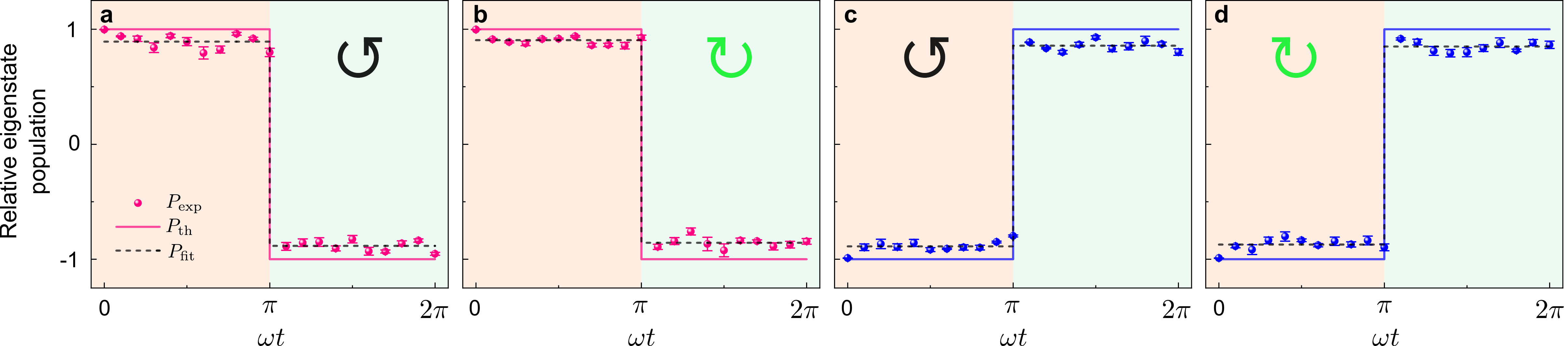}
\caption{{\bf Experimental observation of adiabatic state transfer while dynamically encircling around an EP.} The circle in the inset  represents the counterclockwise or clockwise encircling direction in the $(x,y)$-plane defined in Eq.~(\ref{x-y}), as shown in Fig.~\ref{figure1}({\bf a}). {\bf a-b} The initial state is chosen as $|\psi_1\rangle$ on the $E_1$ energy surface. {\bf c-d} The initial state is chosen as $|\psi_2\rangle$ on the $E_2$ energy surface.
The region from $0$ to $\pi$ along the horizontal axis represents uncompleted state switching for times $0\leq t\leq T/2$, and the region from $\pi$ to $2\pi$ along the horizontal axis, corresponding to times $T/2\leq t\leq T$, represents successful state switching. The solid circles represent the experimental data and the solid lines denote the theoretical prediction. Dashed curves are obtained by numerically fitting the experimental data, where the fitting values obtained for the uncompleted/completed state switching region of each panel are: ($\mathbf{a}$) 0.89/-0.88; ($\mathbf{b}$) 0.90/-0.86; ($\mathbf{c}$) -0.89/0.86; ($\mathbf{d}$) -0.87/0.85.
Error bars represent the standard deviation of the measurement results. The discrepancy between experiment and theory mainly stems from the uncertainty of the instantaneous eigenstates and unavoidable noise during the experiment. The other parameters are $r=0.5$ and $\phi_0=\pi$.}
    \label{figure2}
\end{figure*}

\addtocontents{toc}{\protect\setcounter{tocdepth}{-10}}

\section*{Results}
\subsection*{Adiabatic and Symmetric State Transfer in Time-Modulated Non-Hermitian Systems}In a properly chosen orthonormal basis denoted by $\{|0\rangle, |1\rangle\}$, we consider a two-level non-Hermitian Hamiltonian (NHH)~\cite{AMI+24}
\begin{eqnarray}\label{H_0}
    H_0 = \begin{pmatrix}
        -iv+\epsilon & u+i\kappa \\
        u+i\kappa & iv-\epsilon
    \end{pmatrix},
\end{eqnarray}
where all parameters in Eq.~\eqref{H_0} are assumed to be real, and the non-Hermiticity is introduced by the parameters $v$ and $\kappa$ ~\cite{CLW+22}.
Accordingly, the NHH governs the state evolution via the Schr\"odinger equation (in units $\hbar=1$)
   $ i\partial_t|{\psi}(t)\rangle=H_0|\psi(t)\rangle$.

The spectrum of the NHH reads
\begin{eqnarray}\label{Epm}
    E_{\pm} = \pm \sqrt{(u-v +i\kappa -i\epsilon) (u+v +
i\kappa +i\epsilon)}\in {\mathbb C},
\end{eqnarray}
 which exhibits EPs when $\{ v_{\rm EP} =\pm u, \epsilon_{\rm EP} = \pm \kappa \}$~\cite{AMI+24}.
When $\kappa=0$, the NHH in Eq.~\eqref{H_0} reduces to the standard NHH known to exhibit chiral state transfer while dynamically encircling the EP~\cite{Doppler2016}.
This chiral behavior originates from the complex spectrum of the NHH: as the eigenstates evolve, the system undergoes NATs that drive it toward least-loss state with the maximal imaginary part of the eigenvalue.
Consequently, the system preferentially evolves into this least-loss state, thus resulting in the observed chirality~\cite{Ozdemir2019}.
This mechanism further suggests that if an NHH could be engineered to host EPs while maintaining a purely real spectrum, NATs would be suppressed; thus enabling the long-sought realization of adiabatic and symmetric state transfer that exploits the unique EP topology.

A theoretical study~\cite{AMI+24} has shown that such adiabatic and therefore symmetric state transfer,
in two-level NHH systems, can indeed be restored when the system parameters are constrained to a hyperboloid submanifold of the parameter space. An extension of this protocol to multi-level systems is also discussed in Ref.~\cite{arkhipov2025}.
In particular, by employing the parametrization:
\begin{eqnarray}\label{parameters}
    &v = \alpha\sinh\phi_i\sin\phi_r, \quad \epsilon =\alpha\cosh\phi_i\cos\phi_r,& \nonumber \\
    &u = \alpha\cosh\phi_i\sin\phi_r, \quad \kappa =\alpha\sinh\phi_i\cos\phi_r,&
\end{eqnarray}
with $\phi=\phi_r+i\phi_i=\arctan(x+iy)^{-1}\in{\mathbb C}$ and $\alpha=x\sinh\phi_i/\sin\phi_r\in{\mathbb R}$,
one transforms the original NHH with complex spectrum into NHH
\begin{eqnarray}\label{H}
   H_0\to H = \alpha\begin{pmatrix}
        \cos\phi & \sin\phi \\
        \sin\phi & -\cos\phi
    \end{pmatrix},
\end{eqnarray}
which is characterized exclusively by real eigenvalues
\begin{eqnarray}\label{E12}
    E_{1,2}=\mp\alpha,
\end{eqnarray}
in the entire $(x,y)$ plane, and where we arrange the eigenvalues in descending order, namely $E_1\geq E_2$. The explicit form of the eigenvectors of the NHH $H$ are given in Methods.

Specifically, the system parameters in Eq.~(\ref{parameters}) define a two-dimensional (2D) hyperboloid embedded in the 4D parameter space $(\epsilon,u,v,\kappa)$ satisfying $\epsilon^2+u^2-v^2-\kappa^2=\alpha^2$ (see also Supplementary Information for an intuitive geometric visualization of this manifold).
Importantly, by restricting the system parameters to such a 2D manifold within the full 4D parameter space, one can effectively reduce the NHH $H_0$ in Eq.~(\ref{H_0}), which generally exhibits complex eigenvalues, into a parameter regime where the spectrum becomes purely real. Physically, this restriction enforces a balance between gain and loss that removes the imaginary part of the spectrum.

Owing to the hyperbolic embedding defined by Eq.~(\ref{parameters}), each point $(x,y)$ uniquely determines a single point in $(\epsilon,u,v,\kappa)$ in the full parameter space, and hence a unique Hamiltonian. Therefore, the spectrum of the NHH $H$ is well defined.
More precisely, the representation of this hyperbolic mapping via Eq.~\eqref{H} is one-to-one and continuous throughout the parameter space except at an EP, where the parametrization becomes singular due to the defectiveness of the Hamiltonian $H_0$. This singularity, however, reflects a breakdown of the coordinate representation rather than of the Hamiltonian itself, which remains well-defined in the original parametrization (see below).

The corresponding EPs, in the spectrum of the NHH $H$, appear as branch points of two intersected real-valued Riemann sheets located at $(x=0,y=\pm 1)$. Notably, the eigenvalues of the NHH $H$ remain well-defined at the EP ($\alpha_{EP}=0$), while the eigenvectors given in Eqs.~(\ref{eigen_right}) and (\ref{eign_left}) become singular ($\phi_{EP}\to -i\infty$). This behavior is a direct manifestation of eigenvector coalescence in the original representation of the NHH $H_0$ in Eq.~(\ref{H_0}) (see Supplementary Information for details). We emphasize that, in our protocol, the EP is only encircled and never crossed. Consequently, the eigenvectors of $H$ and their evolution along the winding trajectories remain well-defined throughout the process.

Interestingly, the branch cut $(x=0,-1<y<1)$ coincides with a diabolic line defined by the condition $E_{1,2}=0$ in Eq.~(\ref{E12}). As a result, the EP-encircling protocol with restored adiabaticity considered here, illustrated in Fig.~\ref{figure1}a and discussed in detail below, necessarily involves traversing this diabolic degeneracy. This observation motivates a clarification of the notion of adiabaticity used in the present protocol. In particular, adiabaticity in the Berry--Kato sense, which is based on the following instantaneous eigenstate in the presence of a finite spectral gap, is not directly applicable here. The term ``adiabatic'' we use is in the operational sense of the absence of discontinuous non-adiabatic transitions. This situation differs from conventional EP-encircling schemes in systems with complex spectra, where branch cuts do not generally coincide with extended real-energy degeneracies and the assumptions underlying the standard adiabatic theorem may still be fulfilled. Here, ``adiabatic'' therefore refers exclusively to a dynamical regime in which sufficiently slow parameter modulation yields continuous state evolution, with adiabaticity identified operationally through the absence of NATs. This dynamical behavior is enabled by the absence of an imaginary component in the spectrum of the NHH $H$ along the encircling trajectory, which suppresses mode-selective amplification and thereby prevents discontinuous NATs~\cite{Doppler2016}. In what follows, unless explicitly stated otherwise, we use the term adiabaticity exclusively in this operational sense.

In this respect, the symmetric state-switching protocol based on dynamical EP encirclement in such systems with a purely real spectrum echoes the adiabatic rapid passage (ARP) protocol familiar from Hermitian systems, where robust state transfer can be similarly achieved through slow parameter variation by traversing diabolic degeneracies during the state evolution~\cite{Metcalf2017}. Similar parallels between these Hermitian and non-Hermitian state-flipping protocols have already been discussed and experimentally explored in Ref.~\cite{Feilhauer2020}.
Crucially, however, compared to Ref.~\cite{Feilhauer2020}, the protocol experimentally implemented here operates in a parameter regime where the diabolic line emerges as the branch cut terminating at EPs, and where the continuous state evolution remains entirely free of NATs.

\subsection*{Experimental simulation of adiabatic and symmetric  state transfer}
We now demonstrate that the dynamics governed by the Hamiltonian in Eq.~(\ref{H}) is independent of NATs, implying that the system can evolve adiabatically along an orbit in parameter space while dynamically encircling an EP. We first choose the time-evolution trajectory as
\begin{eqnarray}\label{x-y}
    x(t) &=& r\sin(\omega t+\phi_0), \nonumber \\
    y(t) &=& 1-r\cos(\omega t+\phi_0),
\end{eqnarray}
where $r,\omega,\phi_0$ are real constants.
Correspondingly, we change $H(t)$ in Eq.~(\ref{H}) along with $x(t)$ and $y(t)$.
This path traces a circle with a radius $r$ on the plane $(x,y)$, centered at the EP, $(x,y)_{\rm EP}=(0,1)$. The starting point corresponds to the phase $\phi_0=\pi$, where the two energy levels $E_{1,2}$ are maximally separated. For angular frequencies $\omega>0$ $(\omega<0)$, the encircling direction is counterclockwise (clockwise).
Note that a loop in the 2D plane $(x,y)$ corresponds to a loop on a 4D hyperboloid in the system parameter space, according to the definition of parameterization (see also Supplementary Information).

In our experiment, we encode the qubit basis in the horizontal and vertical polarization states of a light pulse, namely, $|0\rangle \equiv |H\rangle$ and $|1\rangle \equiv |V\rangle$. The two orthogonal polarization components are coupled to each other via a polarization controller. Figure~\ref{figure1}b shows the experimental optical setup used to simulate the adiabatic dynamics in a non-Hermitian system described by the NHH in Eq.~\eqref{H}. The experimental platform consists of three main stages: state preparation, evolution, and measurement~(see also {\it Methods}).

For a given initial polarization state $|\Psi_0\rangle$, the system evolution is described by
\begin{equation}
|\Psi(t)\rangle = U(t)|\Psi_0\rangle ,
\end{equation}
where $|\Psi(t)\rangle = [a(t), b(t)]^{T}$ represents the polarization state at evolution time $t$, with $a(t)$ and $b(t)$ denoting the horizontal and vertical field components, respectively.

In the experiment, we simulate the continuous system dynamics, generated by a time-dependent NHH $H(t)$,  by constructing a family of `static' non-unitary evolution operators $U(t)$, each of which corresponds to the time evolution from $t=0$ to a chosen target time $t$.
Specifically, in the experiment, we select 21 representative target times that include the initial time. Each target time $t_q$ corresponds to an independently constructed operator $U(t_q)$.
For example, for $\omega t_q = \pi$, the operator $U(t_q)$ represents the evolution from $t=0$ to $t_q=\pi/\omega$.
This operator is then applied to the initial state in a single experimental run (as shown in Fig.~\ref{figure1}).
As such, hereafter, unless stated otherwise, the terms ``time evolution'', ``dynamically encircling'' and ``system dynamics'' refer experimentally to this simulated evolution, rather than to a genuine ``in-situ'' encircling.

To obtain each $U(t_q)$, we first numerically integrate the Schr\"odinger equation from $t=0$ to $t=t_q$. This integration is performed using $N=5\times10^4$ time steps, which ensures the convergence of the time-ordered exponential (see also Supplementary Information).
Formally, the evolution operator is approximated as
\begin{equation}\label{U}
U(t_q) \approx \prod_{n=1}^{N} \exp\!\left[-i H(t_n)\,\delta t\right],
\end{equation}
where $\delta t = t_q/N$ and $t_n=(n-\tfrac{1}{2})\delta t$ denotes the midpoint of each integration interval.
Each such numerically constructed operator $U(t_q)$ is then physically implemented by appropriately configuring the optical elements, thereby directly mapping the integrated evolution onto the output state. Repeating this procedure for different target times $t_q$ allows us to efficiently simulate the continuous dynamics of the system without relying on incremental or stroboscopic parameter modulation.

In the experiment, we implement this sequence of operators using variable polarization controllers and a split-path gain–loss module within a two-mode optical platform (see also Methods).
\color{black}
The non-unitary operator $U(t)$ is experimentally implemented on the basis of the singular value decomposition
The non-unitary operator $U(t)$ is experimentally implemented on the basis of the singular value decomposition
$
U(t) = R_1 L R_2
$,
where $t$ represents the time parameter, which refers to a specific moment corresponding to the point on the trajectory of the parameters that we select in our experiment.
The operators $R_1$ and $R_2$ are unitary matrices which can be readily realized using a set of wave plates with two quarter-wave plates and one half-wave plate at rotatable angles~\cite{XWZ+19}. The operator $L$ is a real positive-definite diagonal matrix in the following form
$
L = {\rm diag}[l_1,l_2]
$.
The experimental parameters varied for each data point are the rotation angles in the two variable polarization controllers to implement two unitary operators, and the gain/loss ratios related to the attenuation of the variable optical attenuator and the gain of the erbium-doped fiber amplifier to implement the operator $L$. These are derived explicitly from the singular value decomposition of $U(t)$, which is calculated using the parameters of the system's Hamiltonian. This allows us to reconstruct the state trajectory with high fidelity, avoiding the cumulative noise of long-loop light evolution associated with a continuous implementation.

\begin{figure*}[t]
    \includegraphics[width=0.99\textwidth]{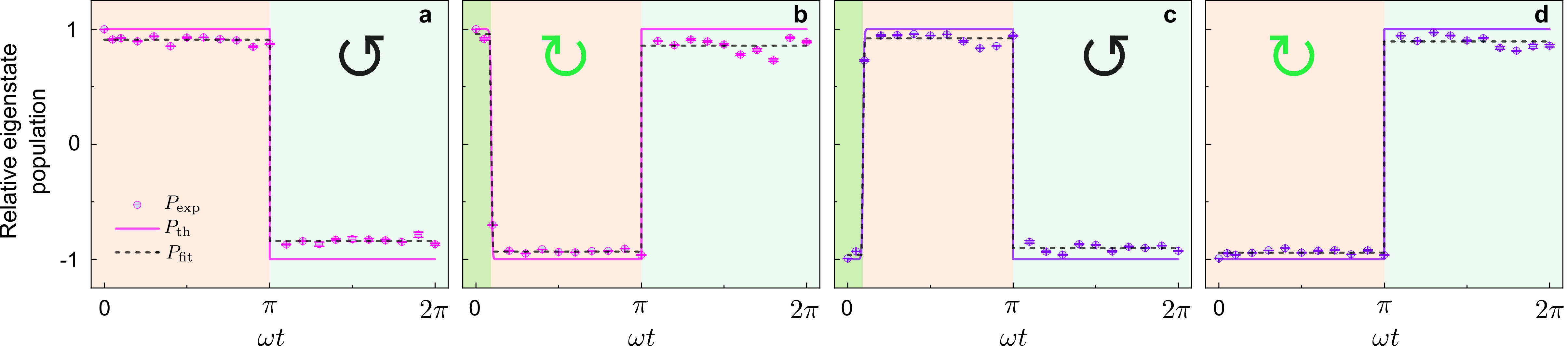}
\caption{{\bf Experimental observation of chiral state transfer while dynamically encircling around an EP.}
{\bf a-b} The initial state is chosen as $|\psi'_1\rangle$. {\bf c-d} The initial state is chosen as $|\psi'_2\rangle$. The circle in the inset  represents the counterclockwise or clockwise encircling direction in the $(x,y)$-plane defined in Eq.~(\ref{x-y}), and the corresponding trajectory in the Riemann surface is shown in the Supplementary Information.
The moment when the relative eigenstate population switches between $+1$ and $-1$ represents the occurrence of the state transfer. {\bf b-c} The non-adiabatic jumps occur in the first interface near the starting point. The other descriptions are the same as those in Fig.~\ref{figure2}.}
    \label{figure2chiral}
\end{figure*}

We choose the initial state of the system as one of the right eigenstates of the NHH $H$, namely, $|\Psi_0\rangle = |\psi_k(t=0)\rangle$ $(k\!=\!1,2)$.
To track the state dynamics, we measure the relative eigenstate population~\cite{NLL+22}
\begin{equation}
    P(t) = \frac{|\langle \eta_1(t) |\Psi(t) \rangle |^2-|\langle \eta_2(t) |\Psi(t) \rangle |^2}{\sum_{k=1,2} |\langle \eta_k(t) |\Psi(t) \rangle |^2},
\end{equation}
where $|\langle \eta_k(t)|\Psi(t)\rangle|^2$ is the fidelity between the evolving state $|\Psi(t)\rangle$ and the instantaneous left eigenstates $|\eta_k(t)\rangle$ of the NHH.
The relative eigenstate population can take values within the interval $P\in[-1,1]$. When $P$ approaches $1$, the evolving state almost entirely populates the first eigenstate $|\psi_1(t)\rangle$. Conversely, as $P$ approaches $-1$, it converges to the second eigenstate $|\psi_2(t)\rangle$.
Here, ``instantaneous'' means that the eigenvectors of the NHH are calculated independently at each time step, for the corresponding values of the parameters $(x,y)$.

At the beginning, we select the initial state as $|\psi_1\rangle$ on the upper energy surface of Fig.~\ref{figure1}a.
As illustrated in Fig.~\ref{figure2}a, in the experiment when the EP is encircled counterclockwise within the time $0\leq t\leq T/2$, the relative eigenstate population $P_{\text{exp}}(t)$ always approaches $1$, indicating that the evolving state $|\Psi_\text{exp}(t)\rangle$, governed by the NHH, is always close to in the first eigenstate $|\psi_1(t)\rangle$. Meanwhile, the state fidelity $|\langle \psi_1(t)|\Psi_{\text{exp}}(t)\rangle|^2$ in this region exceeds $92.91\%$.

In contrast, during the second half of the cycle when $T/2\leq t\leq T$ in Fig.~\ref{figure2}a, the relative eigenstate population $P_{\text {exp}}(t)$ approaches $-1$ and the fidelity between $|\Psi_\text{exp}(t)\rangle$ and $|\psi_2(t)\rangle$ is larger than $94.38\%$,
indicating that the system state successfully switches from $|\psi_1\rangle$ to $|\psi_2\rangle$.
These results imply that after completing the full cycle around the EP, the eigenstates undergo an exchange from $|\psi_1\rangle$ to $|\psi_2\rangle$.
When reversing the encirclement direction to clockwise in Fig.~\ref{figure2}b, the state dynamics exhibits a similar behavior: the initial
eigenstate $|\psi_1\rangle$ is successfully switched to the eigenstate $|\psi_2\rangle$ at the end of the entire loop around the EP. This confirms that the state transfer is independent of the direction of the encirclement, thus demonstrating the adiabatic nature of the state evolution.

When we consider changing the initial state to $|\psi_2\rangle$ on the lower energy surface, a similar phenomenon is observed as demonstrated in Figs.~\ref{figure2}c,d where the initial eigenstate $|\psi_2\rangle$ is successfully switched to the eigenstate $|\psi_1\rangle$ at the end of the entire loop around the EP. This result also indicates the occurrence of adiabatic state transfer.
We note that the apparent abrupt change of the projection of the evolving state on the instantaneous eigenstates at $t=T/2$ in Fig.~\ref{figure2}, corresponding to the crossing of the diabolic line, arises from the interchange of the instantaneous eigenstates at this degeneracy, in accordance with the adopted labeling convention  $E_1\geq E_2$. Importantly, the state evolution remains fully continuous across this degeneracy, as further illustrated by complementary fidelity plots between the evolving and initial states provided in the Supplementary Information.
The observed continuous adiabatic behavior thus stands in stark contrast to previously studied non-Hermitian systems with EPs, where the presence of complex Riemann surfaces leads to non-adiabatic transitions during state evolution.

Importantly, the symmetric state-switching protocol demonstrated here is not restricted to specific winding periods $T$ or particular EP-encircling trajectory shapes in the $(x,y)$-space~\cite{arkhipov2025}. Instead, it remains effective for a broad range of system parameters in Eq.~(\ref{x-y}), provided that the system evolution is sufficiently slow to suppress fast diabatic effects (see also Supplementary Information). This behavior differs from the scenario considered in Ref.~\cite{Feilhauer2020}, where maintaining the adiabatic character of the EP-encircling protocol requires additional constraints on the system parameters.

We further emphasize that the observed adiabatic symmetric state transfer remains robust even when the NHH $H$ in Eq.~\eqref{H} is weakly perturbed so that its spectrum becomes complex, provided that the perturbations are sufficiently small to prevent the onset of NATs. Additional details and supporting analysis are provided in the Supplementary Information.

\subsection*{Experimental simulation of chiral state transfer}
To confirm that the above parameter modulations indeed restore adiabaticity, we further present an experimental analysis of chiral state transfer realized in the same optical setup.
For this purpose, we introduce another NHH $H'$ that can directly induce the chiral state-switching behavior in the system's dynamics, namely
\begin{eqnarray}\label{H2}
H' = C \big[(i x - y)\hat{\sigma}_z + i\hat{\sigma}_x \big],
\end{eqnarray}
where $C = \sqrt{y_0^2 - 1}\,\sin\!\big(\text{arctanh}(y_0^{-1})\big)$ is a scaling factor with $y_0 = \mathrm{const}$, and the system parameters $(x,y)$ are time-modulated in the same manner as in Eq.~\eqref{x-y}.
This particular choice of $C$ ensures that at $t=0$, when $(x=0, y=y_0)$, the NHH $H$ in Eq.~\eqref{H} and the NHH $H'$ in Eq.~\eqref{H2} coincide, namely, they possess identical eigenspectra at the initial time. Moreover, the real spectrum of $H'$ is similar to that of $H$, including the EP at the same location $(x=0,y=1)$. Importantly, the spectrum of $H'$, compared to $H$, exhibits the imaginary part, which plays the key role in inducing the NATs and therefore the chiral mode behavior occurs~(see Supplementary Information for details).
Thus, by implementing both NHHs, $H$ and $H'$, using the same experimental setup enables a controlled switching between adiabatic (symmetric) and non-adiabatic (chiral) mode-transfer regimes for the same set of eigenstates, as we demonstrate below.

We also implemented an experiment analogous to that described in the previous section on adiabatic state transfer. Here, we select the winding speed of $\omega=\pi/500$.
As illustrated in Fig.~\ref{figure2chiral}, starting from different initial states $|\psi'_k\rangle$ (identical to $|\psi_k\rangle$ at $t=0$), a complete counterclockwise encirclement of the EP drives the relative eigenstate population $P_{\text{exp}}(t)$ toward $-1$, indicating that the final state always evolves into $|\psi'_2\rangle$, as shown in Figs.~\ref{figure2chiral}a,c.
Conversely, clockwise encirclement always results in the final state $|\psi'_1\rangle$, we can see the final relative eigenstate population $P_{\text{exp}}(t)$ always approaching $1$ in Figs.~\ref{figure2chiral}b,d.
This experimental observation thus confirms the established results on chiral mode switching in a two-level non-Hermitian system while dynamically encircling an EP~\cite{Ozdemir2019}. This chirality arises because of NATs due to the nonzero difference in the imaginary parts of the two eigenenergies~\cite{Doppler2016}.  Contrary to the adiabatic dynamics, the observed asymmetry indicates that, in this case, the encircling direction rather than the initial conditions determines the final state.

\section*{Discussion}

Although the protocol presented here has been implemented using a classical optical platform, its extension to the quantum regime, for instance, using single photons, is in principle feasible~\cite{Tang2023,GSQ+25} and represents a promising direction toward the realization of robust quantum gates. In the classical setting with coherent light, optical losses can be compensated by amplifying elements, which may also provide controlled gain along the propagation path. In contrast, in the quantum regime, linear amplification unavoidably introduces quantum noise, which degrades coherence and fidelity of fragile quantum states~\cite{Scheel2018}. To circumvent this limitation, relative gain can instead be implemented using purely passive attenuation, thereby avoiding noise injection~\cite{Klauck2019}. This approach is inherently probabilistic, as not all photons survive the evolution, and consequently requires post-selection to discard loss events. While such a passive strategy increases overall losses and limits practical scalability, it nevertheless provides a viable route for experimentally accessing the quantum regime without compromising state coherence.

The non-Hermitian state transfer protocol implemented here can also be understood as a natural extension of the ARP protocol known in Hermitian systems~\cite{Metcalf2017,Feilhauer2020}. An important conceptual distinction between the two is the role played by dissipation. In Hermitian settings, dissipation and decoherence are typically detrimental and must be minimized to preserve adiabatic state transfer. By contrast, in the non-Hermitian framework experimentally realized here, dissipation is not merely assumed but directly exploited as a control resource. Specifically, the engineered non-Hermitian dynamics give rise to parameter-space trajectories along which the spectrum of the effective Hamiltonian becomes purely real, thus suppressing mode-selective amplification and preventing discontinuous NATs during the EP encirclement. As a result, dissipation plays a constructive role in stabilizing the adiabatic dynamical state evolution and enabling symmetric, direction-independent state transfer. Our experimental findings thus further highlight how non-Hermitian control can transform dissipation from a limiting factor into a functional element for stable state manipulation in real-world dissipative photonic architectures.

In this work, we report experimental evidence that adiabatic state evolution is fully restored during an EP-encircling protocol, which is realized through a sequence of individual evolution operators that scan the parameter space.
This is achieved by selecting trajectories in parameter space along which the evolution operator attains a real spectrum. Using a two-mode optical platform, we realize adiabatic evolution by manipulating the polarization state of a laser pulse through precisely controlled intensity and phase modulation operations.

Furthermore, we experimentally demonstrate the ability to switch between symmetric (adiabatic) and asymmetric (non-adiabatic) state transfer regimes for the same set of initial eigenmodes by appropriately varying the system parameters. In other words, we realize a fully programmable symmetric–asymmetric two-mode switch which has been thought impossible to implement in such two-level systems.  This is in striking contrast to previous studies~\cite{arkhipov2023}, where the realization of a programmable state switch requires high-dimensional multimode systems and relies on the nontrivial interplay between exceptional and diabolic points in the Hamiltonian spectrum.

These results open new avenues for non-Hermitian state-control protocols in both classical and quantum photonic platforms, with potential implications for topological quantum computation~\cite{YQSW+21,ZTF+23,QC+14,QC+23,Zhang2018_hol} and related areas of classical and quantum information processing. In particular, the symmetric state-switching protocol demonstrated here in a minimal two-level photonic building block can, in principle, be scaled to implement controlled topological state permutations in multi-qubit systems~\cite{arkhipov2025}; for example, in integrated photonic architectures or networks of coupled waveguides and/or cavities, where it could realize SWAP-type quantum gate operations~\cite{Cheng2023}. The controlled implementation of symmetric group operations in the non-Hermitian qubit eigenspace may enable topological quantum-computation schemes, where resulting holonomies generated by encircling EPs can be exploited as robust logical operations~\cite{Zanardi1999,Duan2001,ZTF+23}.

\section*{Methods}
\subsection*{Eigenvectors of the NHH $H$}
The eigenvectors of the NHH $H$ in Eq.~(\ref{H}) can be readily derived as
\begin{equation}\label{eigen_right}
   |\psi_1\rangle=
        \left[-\sin\frac{\phi}{2},
        \cos\frac{\phi}{2}\right]^T, \quad
    |\psi_2\rangle=
        \left[\cos\dfrac{\phi}{2},
        \sin\dfrac{\phi}{2}\right]^T,
\end{equation}
and the corresponding left eigenvectors
\begin{equation}\label{eign_left}
 |\eta_1\rangle=
        \left[-\sin\dfrac{\phi^*}{2},
        \cos\dfrac{\phi^*}{2}\right]^T, \quad
    |\eta_2\rangle=
        \left[\cos\dfrac{\phi^*}{2},
        \sin\dfrac{\phi^*}{2}\right]^T,
\end{equation}
where $T$ stands for transpose, and the right eigenvectors and corresponding left
eigenvectors form the biorthogonal basis~\cite{B14,QCW+23}, namely, $\langle\eta_j|\psi_k\rangle=\delta_{jk}$.

The right eigenvectors satisfy
$H|\psi_k\rangle = E_k|\psi_k\rangle$, and the corresponding left eigenvectors $|\eta _j\rangle $ are defined via the equation $H^{\dagger}|\eta _j\rangle =E_j^{*}|\eta _j\rangle $.
The set of left and right eigenvectors forms the biorthonormal basis
\begin{equation}
\langle\eta_j|\psi_k\rangle=\delta_{jk},
\end{equation}
along with the closure relation
\begin{equation}
\sum_j |\eta_j\rangle\langle\psi_j| = \sum_j |\psi_j\rangle\langle\eta_j| =\emph{I}.
\end{equation}

It is important to note that the right eigenstates alone are, in general, non-orthogonal. The introduction of the left eigenstates, which span the dual vector space, is therefore essential.
The non-orthogonality of right eigenvectors can be interpreted as a manifestation of the ``non-flatnes'' in the NHH-Hilbert space.
Such a curved space must be endowed with a metric tensor $\nu$, where the left eigenvectors enable us to define a consistent tensor $\nu = \sum |\eta_m\rangle \langle \eta _m|$. The metric $\nu$ is essential for properly defining inner products between arbitrary (right) vectors in this non-flat space, $|\psi _{m,n}\rangle $ as $\langle \psi_m|\psi _n\rangle _{{\rm flat}}\to \langle \psi _m|\nu|\psi _n\rangle_{{\rm curved}}$, ensuring norm preservation and probabilistic interpretation throughout the time evolution~\cite{AMI+24,CAG+19,M10}.

\subsection*{Experimental implementation: state preparation,  evolution and measurement procedure}

In our experiment, as mentioned above, we
simulate the continuous system dynamics
by appropriately constructing the corresponding operator $U(t)$, which effectively mimics the induced time evolution.
This is achieved as follows.  At the beginning, by restricting the system parameters to the 2D manifold within the full 4D parameter space, one  reduces the NHH $H_0(t)\rightarrow H(t)$, which generally exhibits complex eigenvalues, into a parameter regime where the spectrum becomes purely real. For a given point on the trajectory, the Hamiltonian parameters $\{ v(t),\epsilon(t), \kappa(t), u(t)\}$ are determined by $\{x(t),y(t)\}$.
After that, we  compute the target cumulative evolution operator $U(t)$, according to Eq.~(\ref{U}), and decompose this target operator into the form $U(t) = R_1 LR_2$ via the singular value decomposition.

In practice, these four parameters $\{ v(t),\epsilon(t), \kappa(t), u(t)\}$ collectively contribute to each element of the matrix $U(t)$. However, for coherence parameters (detuning $\epsilon$, coherent coupling $u$), the unitary parts of the evolution, which mix the polarization modes (basis rotations), are mainly encoded in the rotation matrices $R_1$ and $R_2$. These are implemented by adjusting the waveplate angles in the two variable polarization controllers. For non-Hermitian parameters (gain/loss $v$, dissipative coupling $\kappa$), the non-unitary feature is primarily encoded in the diagonal matrix $L$. This is physically implemented by controlling the gain/loss ratio between the two split optical paths using the variable optical attenuator and erbium-doped fiber amplifiers together with the optical delay line.

As illustrated in Fig.~\ref{figure1}b, our experimental platform consists of three main stages: the state preparation, the evolution, and the measurement.

In our work, the initial states we select are two right eigenstates in Eq.~(\ref{eigen_right}). After normalizing each of them, we encode the two orthogonal components of each state onto the horizontal and vertical polarizations of the polarized light, which matches well to the Jones vector~\cite{Jones1941,Hecht2016}. This is conducive to a better experimental implementation. Experimentally, the initial eigenstate can be encoded in the polarization state of the light pulse $\vec{J}=[a_0,b_0]^{T}= |\psi_k(t=0)\rangle$ $(k\!=\!1,2)$, where the state $|\psi_k\rangle$ is normalized. Here, the horizontal and vertical components of a light pulse are represented by $a_0$ and $b_0$, respectively.
As illustrated in Fig.~\ref{figure1}b, a $266.7$ ns laser pulse at a wavelength of $1550$ nm (generated by a distributed feedback laser diode and an acousto-optic modulator) featuring an original state of polarization goes through the variable polarization controller. The variable polarization controller can realize the required $2\times2$ unitary transformation. This is because it is equivalent to a sandwiched combination of two quarter-wave plates and one half-wave plate (Q-H-Q) by controlling the phase delay amount~\cite{Ulrich1979,moroney2022kerr,NLL+22}.
The Jones matrix form for a half-wave plate at the setting angle $\varphi$ is~\cite{GSQ+25}
\begin{eqnarray}\label{Uh}
U_h=\begin{pmatrix}
\cos(2\varphi) & \sin(2\varphi) \\
\sin(2\varphi) & -\cos(2\varphi)
\end{pmatrix},
\end{eqnarray}
and the Jones matrix form for a quarter-wave plate at the setting angle $\varsigma$ is
\begin{eqnarray}
U_q = \begin{pmatrix}\label{Uq}
\cos^2 \varsigma + i \sin^2 \varsigma & (1 - i) \sin \varsigma \cos \varsigma \\
(1 - i) \sin \varsigma \cos \varsigma & \sin^2 \varsigma + i \cos^2 \varsigma
\end{pmatrix}.
\end{eqnarray}
Therefore, by controlling the rotation angle of the tunable polarization controller, we can prepare the desired initial state~\cite{NLL+22}.
After measurement with a polarimeter, we can verify the quality of the state preparation.  The polarimeter we use can obtain the normalized power split ratio and the phase difference of the horizontal and vertical polarization components, which can determine the form of the polarization state. We calculate the fidelity between the experimentally prepared initial states and the theoretically designed initial states as: $99.92\% \pm 0.04\% $ for $|\psi_1\rangle $ and $99.72\% \pm 0.05\% $ for $|\psi_2\rangle $, indicating that the prepared initial states are highly consistent with the theory. Then the pulse with the prepared initial state is injected into the main optical path and subsequent operations are carried out.

After the initial state preparation, by employing two such variable polarization controllers, we realize the unitary matrices $R_1$ and $R_2$.
A polarization beam splitter then separates the horizontal and vertical polarization components of a light beam into two independent modes. Each mode undergoes a distinct gain or loss modulation, enabling the realization of a real diagonal matrix $L$, where $l_1$ and $l_2$ are respectively regarded as the gain and loss coefficients related to an erbium-doped fiber amplifier and the variable optical attenuator.
Additionally, if needed, phase modulation can also be applied to each mode via phase modulators. Here, the optical delay line is also deployed to balance the length of the two modes. Then, the two modes are subsequently recombined using a polarization beam combiner. To compensate for the insertion losses introduced by electro-optical and fiber components, an additional erbium-doped fiber amplifier is inserted into the two-mode rail optical system to prevent pulse attenuation.

At the end of the evolution path, half of the optical pulses are directed to a polarimeter for polarization state analysis. The remaining pulses are used for monitoring purposes via an oscilloscope. In the experiment, we select the discrete time points ${(\omega t)}_m=(m\Delta N-\frac{1}{2})\times\frac{2\pi}{N}$ with $\Delta N=N/20$, $m=1,\dots,20$, $\delta_t=t/N$, $\omega=\pi/50000$ and $N=5\times10^{4}$, as shown in Fig.~\ref{figure2}.
Here, $N$ refers to the number of steps in the numerical integration (the total number of segments as mentioned above) used in the theoretical calculation of the target operator $U(t)$. We use a large $N$ to ensure the theoretical matrix is computed with high precision, rather than feed the signal back $5 \times 10^4$ times in the experiment.
Then, we experimentally measure the polarization state of the system, projected onto the instantaneous left eigenstates, and thereby obtain the relative eigenstate population $P(t)$.
The deviations are dominated by experimental imperfections in implementing the non-unitary evolution operator and measurement uncertainties, which include contributions from instrument instability, environmental fluctuations, and disturbances introduced by the optical fibers.

\vspace{3mm}

{\bf Data availability}

All data are available from the corresponding authors upon reasonable request.

\vspace{5mm}

{\bf Code availability}

The code for simulating the states fidelity is available from the corresponding authors upon reasonable request.

\vspace{5mm}

{\bf Acknowledgments}

This work has been supported by the National Key R\&D Program of China (Grant No.~2023YFA1406701) and National Natural Science Foundation of China (Grant Nos.~92265209, 92476106, 12504587, 12305008, 12374479, 12504413, 12025401). I.A. acknowledges support from the Grant Agency of the Czech Republic (Project No.~25-15775S), and from the Ministry of Education, Youth and Sports of the Czech Republic Grant OP JAC No. CZ.02.01.01/00/23\_021/0008790. F.N. is supported in part by the Japan Science and Technology Agency (JST) [via the CREST Quantum Frontiers program Grant No. JPMJCR24I2, the Quantum Leap Flagship Program (Q-LEAP), and the Moonshot R\&D Grant Number JPMJMS2061]. Q.L. acknowledges support from the China Postdoctoral Science Foundation (Grant No. BX20240065 and No. 2024M750405) and the Basic Research Program of Jiangsu (Grants No. BK20251297). H.X.G. acknowledges support from the China Postdoctoral Science Foundation (Grant No. BX20250174 and No.~2024M760425). D.K.Q. acknowledges support from the China Postdoctoral Science Foundation (Grants No. BX20230036 and No. 2023M730198). L.X. acknowledge support from Beijing National Laboratory for Condensed Matter Physics (No. 2024BNLCMPKF010).

\vspace{5mm}

{\bf Authors contributions }

X.W.W. performed the experiment and interpreted the results with contributions from Q.L., H.X.G., D.K.Q., and L.X.. I.I.A developed the theoretical aspects and interpreted the results with contributions from F.N.. P.X. supervised the project, and analysed the results. All authors contributed in writing the manuscript.

\vspace{5mm}

{\bf Competing interests}

The authors declare no competing interests.

\bibliography{references}

%
%



\restoreTOC
\clearpage %
\onecolumngrid
\begin{center}
\textbf{\large Supplementary Information }
\end{center}
\setcounter{equation}{0}
\renewcommand{\theequation}{S\arabic{equation}}
\renewcommand{\cite}[1]{\citep{#1}}
\renewcommand{\figurename}{\textbf{Supplementary Fig.}}
\renewcommand{\thefigure}{\textbf{\arabic{figure}}}
\setcounter{figure}{0}

\author{Xiaowei Wang}\thanks{These authors contributed equally to this work.}
\affiliation{Beijing Computational Science Research Center, Beijing 100193, China}
\author{Ievgen I. Arkhipov}\thanks{These authors contributed equally to this work.}
\affiliation{Joint Laboratory of Optics of Palack\'y University and Institute of Physics of CAS, Faculty of Science, Palack\'y University, 17. listopadu 12, 771 46 Olomouc, Czech Republic}
\author{Quan Lin}
\author{Huixia Gao}
\author{Dengke Qu}
\author{Lei Xiao}
\affiliation{School of Physics, Southeast University, Nanjing 211189, China}
\author{Franco Nori}
\affiliation{Quantum Information Physics Theory Research Team, Quantum Computing Center, RIKEN, Wakoshi, Saitama, 351-0198, Japan}
\author{Peng Xue}\email{gnep.eux@gmail.com}
\affiliation{School of Physics, Southeast University, Nanjing 211189, China}

\maketitle

\onecolumngrid

\tableofcontents
\clearpage

In this Supplementary Material, we provide details on: encircling loops on the hyperboloid manifold, the spectrum of the non-Hermitian Hamiltonian (NHH) $H'$ and the prospective encircling trajectories around the exceptional point (EP), a proposed experimental scheme for the dynamical observation of the adiabatic state transfer, perturbation analysis, robustness analysis of evolution speed and loop parameters on restored adiabaticity, and additional supporting experimental information.

\subsection{Supplementary Note 1. Encircling Loops on the Hyperboloid Manifold in the 4D Parameter Space}

In the main text, we describe how the encircling loops in the $(x,y)$ plane correspond to loops on the hyperboloid surface in the four-dimensional (4D) parameter space $[\epsilon(x,y),u(x,y),v(x,y),\kappa(x,y)]$, due to the parameterization definition in the main text. In Supplementary Fig.~\ref{figureS4} we illustrate the actual shape of such loops on these hyperboloid surfaces in the 4D space.
\begin{figure}[b]
    \includegraphics[width=0.88\textwidth]{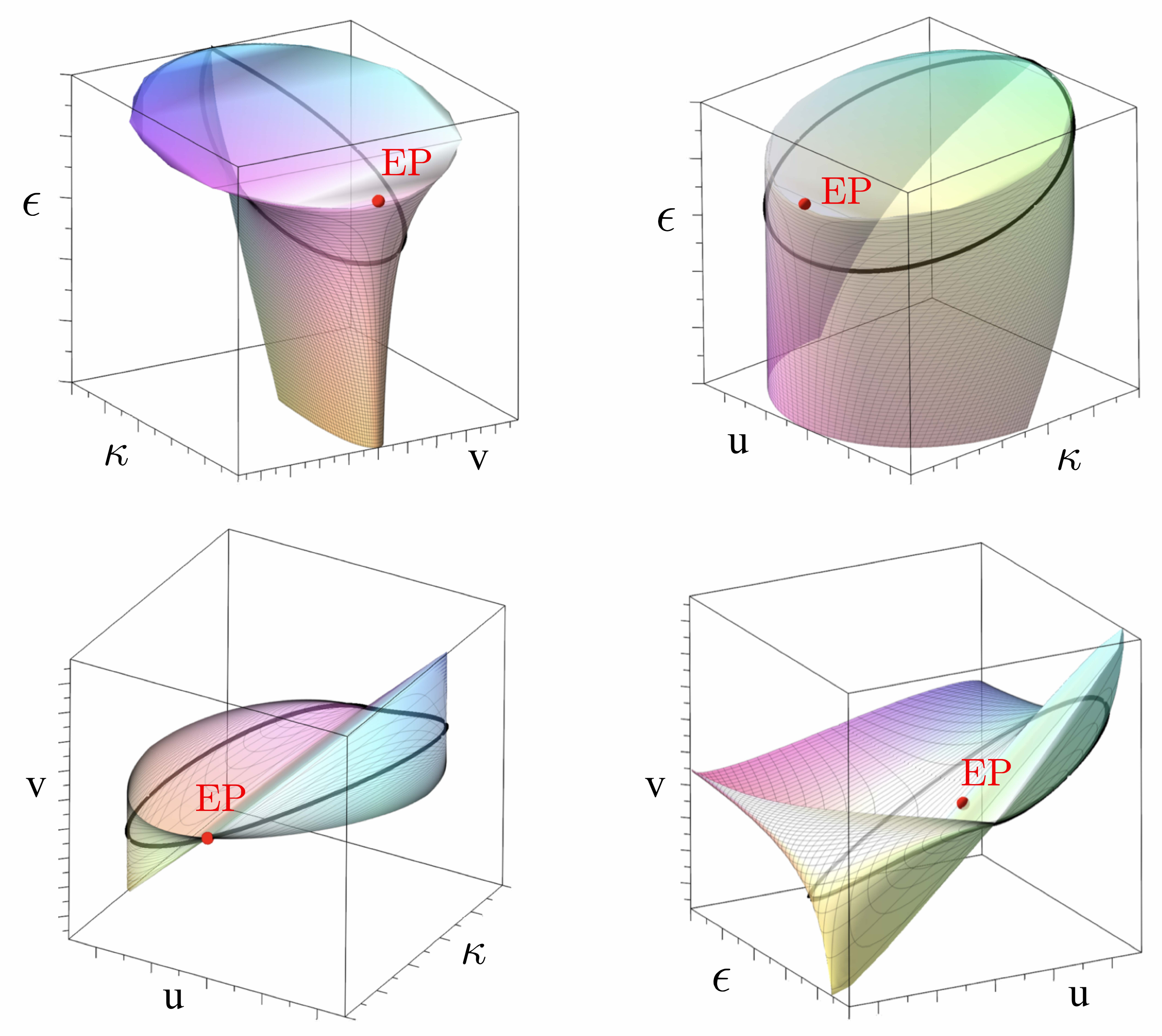}
\caption{ A loop (black solid curve) on the hyperboloid submanifold (shown as a 2D colored surface)  encircling the exceptional point (red dot) in the 4D parameter space $(\epsilon, {u}, {v}, \kappa)$. The winding trajectory in 4D space is illustrated through four different 3D projections. This encircling path corresponds to a circular loop in the $(x,y)$ plane, as defined in Eq.~(3) for $r=0.5$, and shown in Fig.~1a in the main text. }
    \label{figureS4}
\end{figure}

\subsection{Supplementary Note 2. Exceptional Point and Resolution of Parametrization Singularity}
\begin{figure}[t]
    \includegraphics[width=0.88\textwidth]{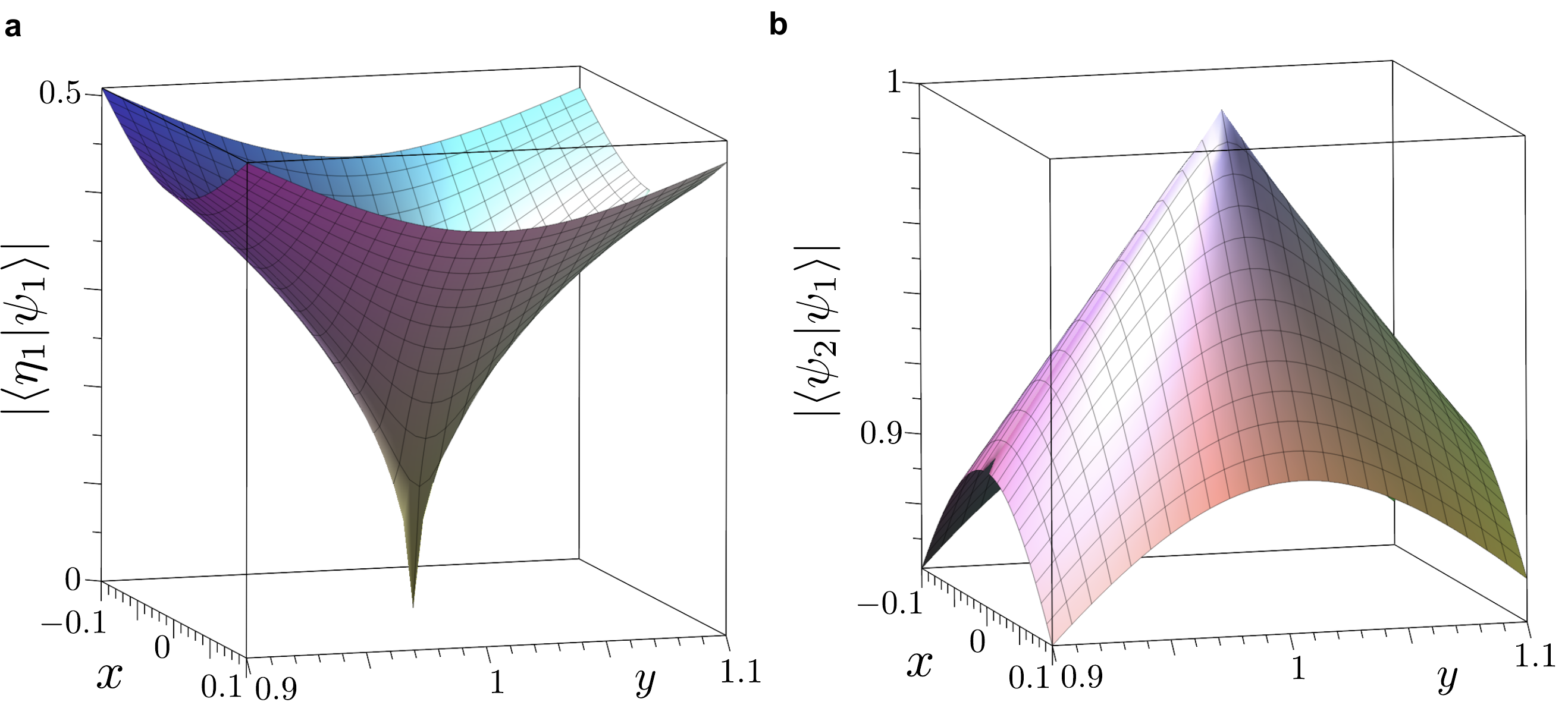}
\caption{Eigenvectors overlap of the non-Hermitian Hamiltonian $H_0$ [Eq.~(1) in the main text], demonstrating the characteristic signatures of an exceptional point (EP) in the $(x,y)$ parameter space.
$\mathbf{a}$ Biorthogonal overlap between left and right eigenvectors, which vanishes at the EP located at $(x=0, y=1)$, indicating self-orthogonality.
$\mathbf{b}$ Overlap between the two right eigenvectors of $H_0$, which approaches unity at the same point, signaling the coalescence of eigenvectors at the EP.}
    \label{figureS11}
\end{figure}
In the main text, we introduced a transformed non-Hermitian Hamiltonian $H_0 \to H$ obtained by mapping the system parameters onto a hyperbolic surface [Eqs.~(3) and (4)]. This mapping is continuous throughout the parameter space, except at the exceptional point (EP), where the original non-Hermitian Hamiltonian $H_0$ becomes defective.

As a consequence, the eigenvectors of the transformed Hamiltonian $H$, given in Eqs.~(11) and (12), exhibit an apparent divergence when approaching the EP, since $\phi\to -i\infty$ at this point. At first sight, this behavior seems inconsistent with the expected properties of an EP, where the right eigenvectors coalesce, $\langle \psi_1^{EP} | \psi_2^{EP} \rangle = 1$, and the biorthogonal norm vanishes, $\langle \eta_i^{EP} | \psi_i^{EP} \rangle = 0$ for $i=1,2$. This raises the question of whether the singularity in the transformed representation corresponds to a genuine EP.

To clarify this issue, it is useful to return to the original representation, where the non-Hermitian Hamiltonian $H_0$ [Eq.~(1) in the main text] is well-defined across the entire parameter space. By tracking the behavior of its right eigenvectors $|\psi\rangle$ ($H_0|\psi\rangle=E|\psi\rangle$) and left eigenvectors $|\eta\rangle$ ($H_0|\eta\rangle=E^*|\eta\rangle$) as functions of the parameters $(x,y)$, one can directly verify the defining features of an EP at these points. More specifically, we first calculate the eigenspectrum of $H_0$ as a function of the parameters $(\epsilon,u,v,\kappa)$, and then apply the hyperbolic map in Eq.~(3) in the main text.

In particular, we find that the point $(x=0, y=1)$ indeed corresponds to an EP. At this point, the left and right eigenvectors become mutually orthogonal, and the right eigenvectors coalesce, as shown in Supplementary Fig.~\ref{figureS11}a,b, respectively. This demonstrates that the EP is well-defined in the original Hamiltonian, and that the apparent singularity arises solely from the parametrization used in the transformed Hamiltonian. Importantly, our protocol relies exclusively on encircling the EP, i.e., on trajectories that remain away from the singular point itself. Along such cyclic paths, the eigenvectors remain well-defined, ensuring the consistency of the description.

\subsection{Supplementary Note 3. Dynamical Mapping via Discretized Unitary Decomposition}
To elucidate the correspondence between the target physical model and our experimental setup in Fig.~1(b) of the main text, we provide a schematic representation of the simulation protocol in Supplementary Fig.~\ref{figureS_sche}. This diagram illustrates how the continuous dynamics of the target system is mapped onto the controllable quantum simulator through state preparation, a single-pass of light through the optical setup, and final measurement.

\begin{figure}[t]
    \includegraphics[width=0.80\textwidth]{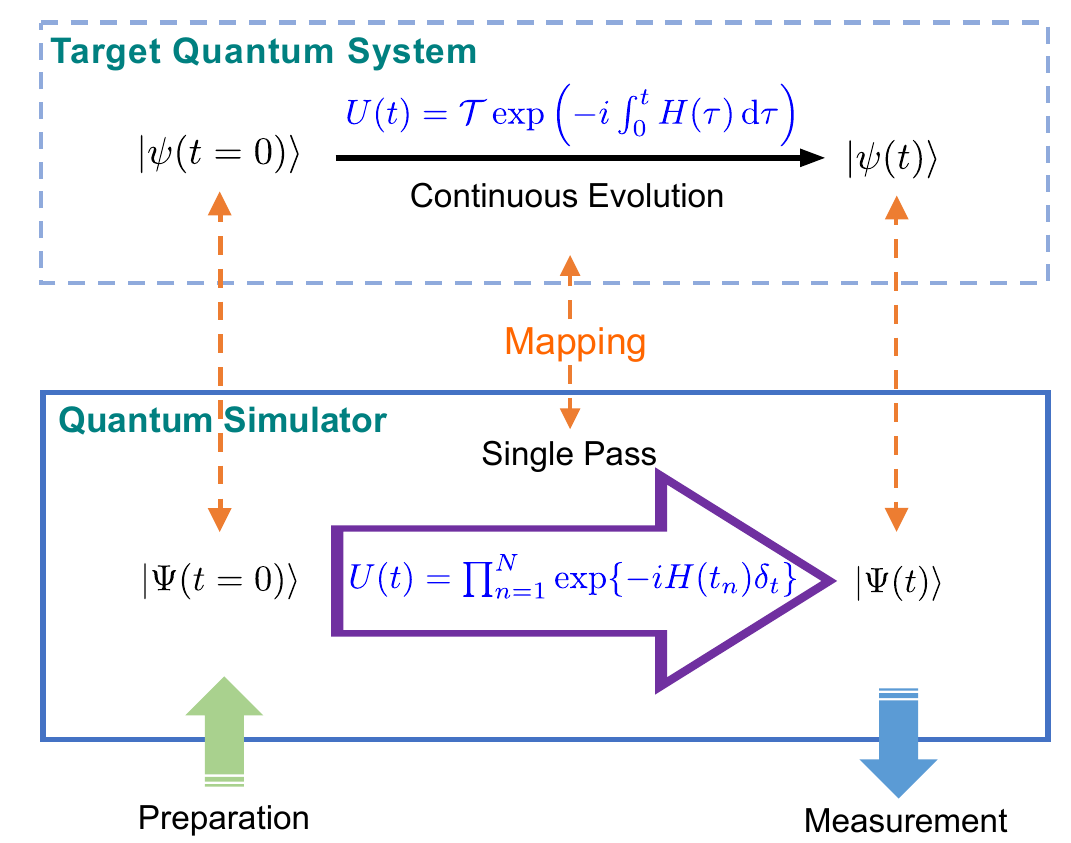}
\caption{ Schematic illustration comparing a target quantum system with its corresponding quantum simulator. The target system (top) evolves from an initial state $|\psi(t=0)\rangle$ to a final state $|\psi(t)\rangle$ under the governing Hamiltonian $H(t)$. The simulator (bottom) replicates this dynamics through a controllable process. Specifically, the simulator is initialized in state $|\Psi(t=0)\rangle$ and driven through a single pass of the discretized unitary evolution $U(t) = \prod_{n=1}^N \text {exp}\{-iH(t_n)\delta_t\}$ to reach $|\Psi(t)\rangle$. The vertical dashed orange arrows indicate the mapping correspondence between the target and simulated states. The upward green arrow and downward blue arrow denote the state preparation and measurement phases, respectively. The detailed experimental implementation can be shown in Fig.~1(b) of the main text. }
    \label{figureS_sche}
\end{figure}

\vspace{12pt}
\subsection{Supplementary Note 4. The Form of States in the Polarization Representation}
In optics, polarized light can be described by the Jones calculus developed by R. C. Jones in 1941~\cite{Jones1941}. Polarized light is represented by Jones vectors, and linear optical elements are represented by Jones matrices.
For an arbitrary polarized light, it can be represented by two orthogonal linear polarization components. Here, we generally select horizontal and vertical polarizations, which can be well separated using a polarizing beam splitter.
Using the Jones matrix notation, the complex amplitudes of the two orthogonal components of the polarized light are:
\begin{eqnarray}
\vec{J}(t) =
\begin{bmatrix} \hat{E}_X \\ \hat{E}_Y \end{bmatrix} =
\begin{bmatrix} E_X e^{i(\omega_0 t - k_0 Z + \delta_X)} \\ E_Y e^{i(\omega_0 t - k_0 Z + \delta_Y)} \end{bmatrix} = \exp\Big({i(\omega_0 t - k_0 Z + \delta_X)}\Big)
\begin{bmatrix} E_X \\ E_Y e^{i\delta_D} \end{bmatrix},
\end{eqnarray}
where $\delta_D=\delta_Y-\delta_X$ represents the phase difference between the two components, $\omega_0$ and $k_0$ denote the  angular frequency and the wavenumber, $E_X$ and $E_Y$ are their amplitudes. By omitting the identical factors, the general expression of the Jones vector after normalization can be written as:
\begin{equation} \label{J}
\vec{J} = \frac{1}{\sqrt{E_{X}^2 + E_{Y}^2}} \begin{pmatrix} E_{X} \\ E_{Y} e^{i\delta_D} \end{pmatrix}.
\end{equation}
Based on the above discussion, we can also use the polarization ellipse and the Poincaré sphere with Stokes parameters to more intuitively describe the polarization of light~\cite{Hecht2016}.

\subsection{Supplementary Note 5. Fidelity Between Evolving and Initial States}
In Fig.~2 in the main text, we present experimental plots of the instantaneous eigenstate population of the NHH in Eq.~(2). Here, in Supplementary Fig.~\ref{figureS_fid}, we provide complementary plots for the fidelity between the evolving and static initial states instead, thus further confirming the observed adiabatic and {\it continuous} symmetric state switching mechanism.

We notice that there is a lower limit for the fidelity in Supplementary Fig.~\ref{figureS_fid}. This theoretical lower bound is determined by the state overlap of the instantaneous eigenstates $|\psi_1(0)\rangle$ and $|\psi_2(0)\rangle$ at the initial time. 
This result is attributed to our intentional protocol design, which aims to maximize the energy gap between the two levels $E_{1,2}$ at time $t=0$; thus ensuring that the initial states have an overlap as minimal as possible, and which equals  $\sim 4/9$.

Compared with Fig.~2 in the main text, we note that the fidelity is significantly more sensitive to noise and experimental imperfections than the population metric $P$ utilized in the main text.

The relative eigenstate population $P$, given in Eq.~(9) in the main text, is calculated within the biorthogonal basis. In non-Hermitian systems, the right eigenstates $\{|\psi_m\rangle\}$ are not mutually orthogonal, but they form a biorthogonal basis with the left eigenstates $\{|\eta_n\rangle\}$, satisfying the condition $\langle \eta_n | \psi_m \rangle = \delta_{mn}$.
To obtain the relative population $P$ experimentally, we project the measured state $|\psi_{\text {exp}}\rangle$ onto the left eigenvectors of the Hamiltonian (i.e., calculating overlaps like $\langle \eta_n | \psi_{\text{exp}} \rangle$). This biorthogonal projection effectively ``filters out'' the contribution from the unwanted mode due to the orthogonality condition $\langle \eta_n | \psi_m \rangle=0$ $(m\neq n)$. Consequently, the population metric $P$ is inherently robust and less sensitive to small imperfections in the evolved state, leading to better agreement between theory and experimental data.

In contrast, the state fidelity, is defined as the direct overlap magnitude
$F=|\langle \psi_{k} | \psi_{\text{exp}} \rangle|^2$
 between the initial and experimentally evolving right states.
Consequently, it becomes highly sensitive to all forms of experimental noise, including global phase fluctuations and intensity drifts. The scatter observed in Supplementary Fig.~\ref{figureS_fid} reflects this hypersensitivity to experimental noise, rather than a fundamental deviation from the theoretical predictions.

\begin{figure}[h]
    \includegraphics[width=0.99\textwidth]{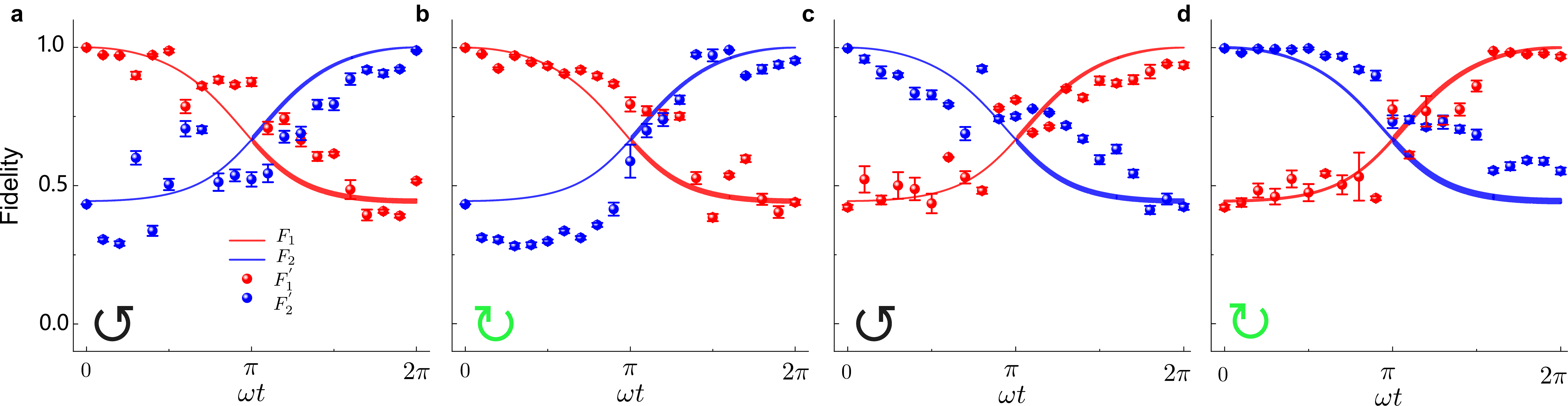}
\caption{Fidelity between the evolving and initial states. This is complementary to Fig.~2 of the main text.
{\bf a-b}: The initial state is chosen as $|\psi_1\rangle$ on the $E_1$ energy surface. {\bf c-d}: The initial state is chosen as $|\psi_2\rangle$ on the $E_2$ energy surface. Fidelity $F_{k}=|\langle \psi_k(t)|\Psi_\text{th}(t)\rangle|^2$ ($k=1,2$) quantifies the overlap between the theoretical time-evolving right eigenstate $|\Psi_\text{th}(t)\rangle$ and instantaneous right eigenvector $|\psi_k(t)\rangle$ of the NHH. Similarly, $F_{k}^{\prime}=|\langle \psi_k(t)|\Psi_\text{exp}(t)\rangle|^2$ ($k=1,2$) denotes the fidelity between the evolving experimental state $|\Psi_\text{exp}(t)\rangle$ and instantaneous right eigenvector $|\psi_k(t)\rangle$ of the NHH while encircling the EP. The state is normalized at each moment, ensuring that the fidelity always lies between $0$ and $1$. The discrepancy between experiment and theory mainly stems from the uncertainty of the instantaneous eigenstates and unavoidable noise during the experiment, as well as the nonlinear quantitative relationship between the fidelity and system states. Other system parameters are identical to Fig.~2.
}
    \label{figureS_fid}
\end{figure}

\subsection{Supplementary Note 6. The Spectrum of the NHH $H'$ and the Encircling Trajectories Around the EP}
As shown in Supplementary Fig.~\ref{figureS_eig}, we plot the energy spectrum of the NHH $H'$ of the main text. Importantly, the spectrum of $H'$, compared to $H$, exhibits the imaginary part, which plays the key role in inducing the NATs. To investigate the state transfer process under such circumstances,
we construct the trajectory of the state evolution over the Riemann surface of the non-Hermitian Hamiltonian as follows:
\begin{eqnarray}\label{E_tra}
\bar{E}(t) = \frac{\sum_{i=\pm} E_i(t)|\langle \eta_i(t) |\psi(t) \rangle |^2}{\sum_{i=\pm} |\langle \eta_i(t) |\psi(t) \rangle |^2},
\end{eqnarray}
where $|\psi(t) \rangle$ is the evolving state governed by the NHH and $|\eta_i(t) \rangle $ is the left eigenvector.

To demonstrate the non-adiabatic nature of dynamically encircling an EP, we show trajectories on real energy surfaces with different encircling directions, starting on both of the Riemann sheets involved (shown as red and blue surfaces), as shown in Supplementary Fig.~\ref{figureS_tra}.  We observe that the state that survives at the end of a loop is independent of the initial state and depends solely on the direction of the loop. Counterclockwise looping always results in a state on the lower-energy plane, while clockwise looping always results in a state on the upper-energy plane, exhibiting chiral characteristics. It is not difficult to find that this chiral mode behavior is caused by non-adiabatic transitions, as shown in Supplementary Figs.~\ref{figureS_tra}b,c. This result is consistent with the results in Fig.~3 of the main text.

\begin{figure}
\includegraphics[width=0.99\textwidth]{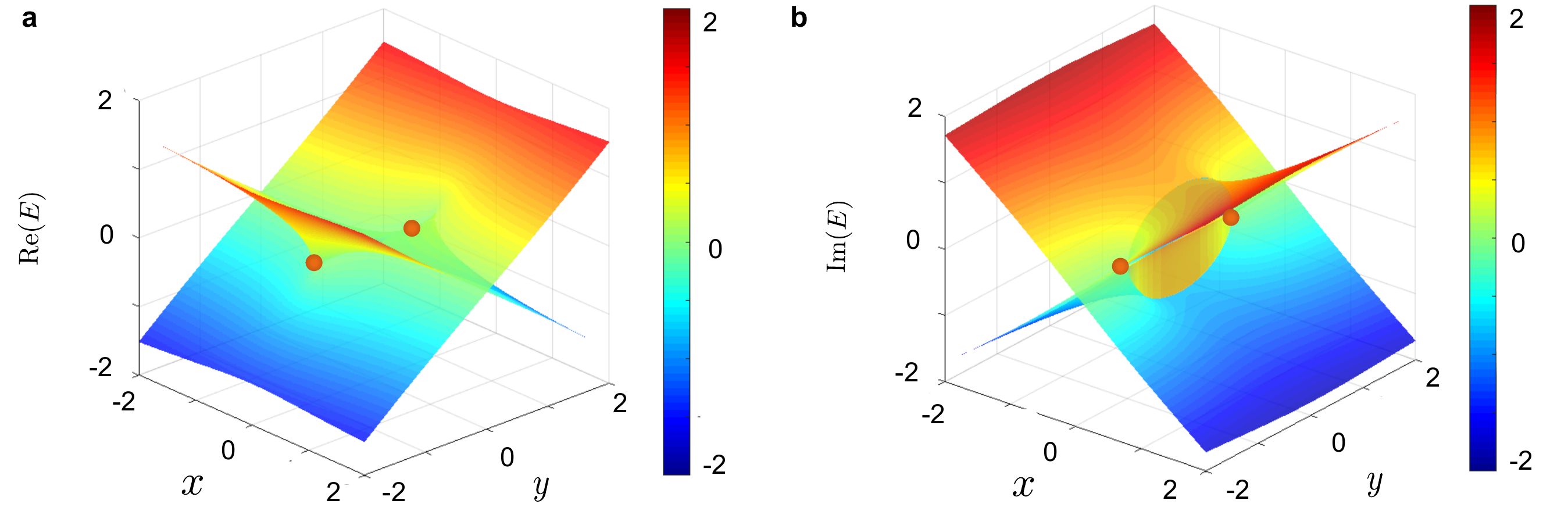}
\caption{The eigenenergy spectrum of the NHH $H'$ in Eq.~(4). $\mathbf{a}$ Real energy spectrum and $\mathbf{b}$ imaginary energy spectrum. The small orange spheres represent the EP points.}
    \label{figureS_eig}
\end{figure}

\begin{figure}
    \includegraphics[width=0.98\textwidth]{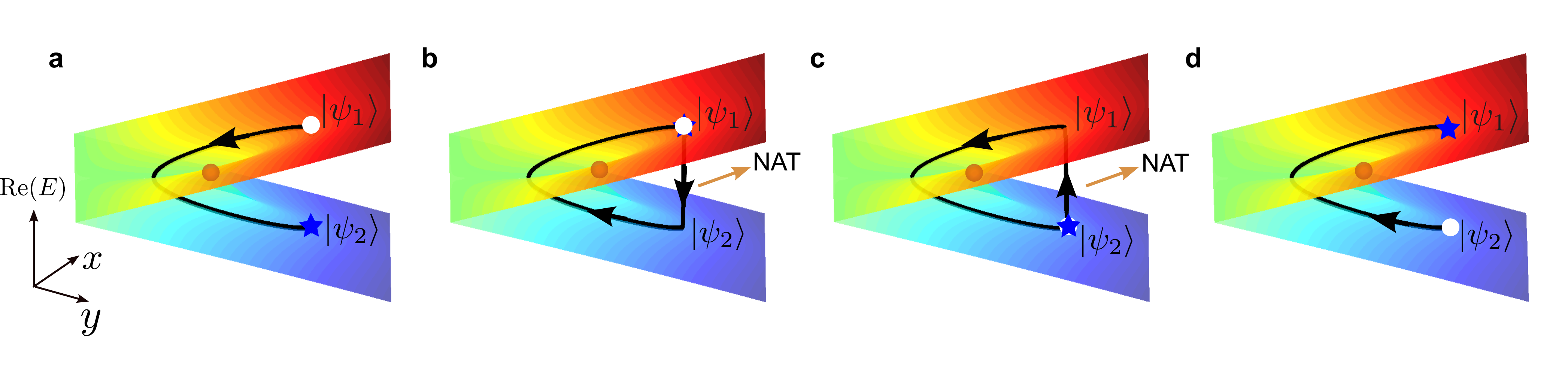}
\caption{Prospective encircling trajectories in the parameter space for the dynamics of two states with starting points on different sheets.
The arrow indicates the encircling direction around the EP. $\mathbf{a},\mathbf{c}$ Counter-clockwise direction. $\mathbf{b}$,$\mathbf{d}$: Clockwise direction.
The starting positions are shown as white circles, and the evolved endpoints are represented by blue stars. {\bf a-b} The initial state is chosen as $|\psi_1\rangle$. {\bf c-d} The initial state is chosen as $|\psi_2\rangle$.  {\bf b-c} The starting point and the ending point coincide, indicating that after one cycle of evolution, it returns to its original state. In this case, non-adiabatic transitions (NATs) occur.}
    \label{figureS_tra}
\end{figure}

\subsection{Supplementary Note 7. A Proposed Experimental Scheme for the Continuous Observation of the Adiabatic State Transfer}

\begin{figure}[b]
   \includegraphics[width=0.95\textwidth]{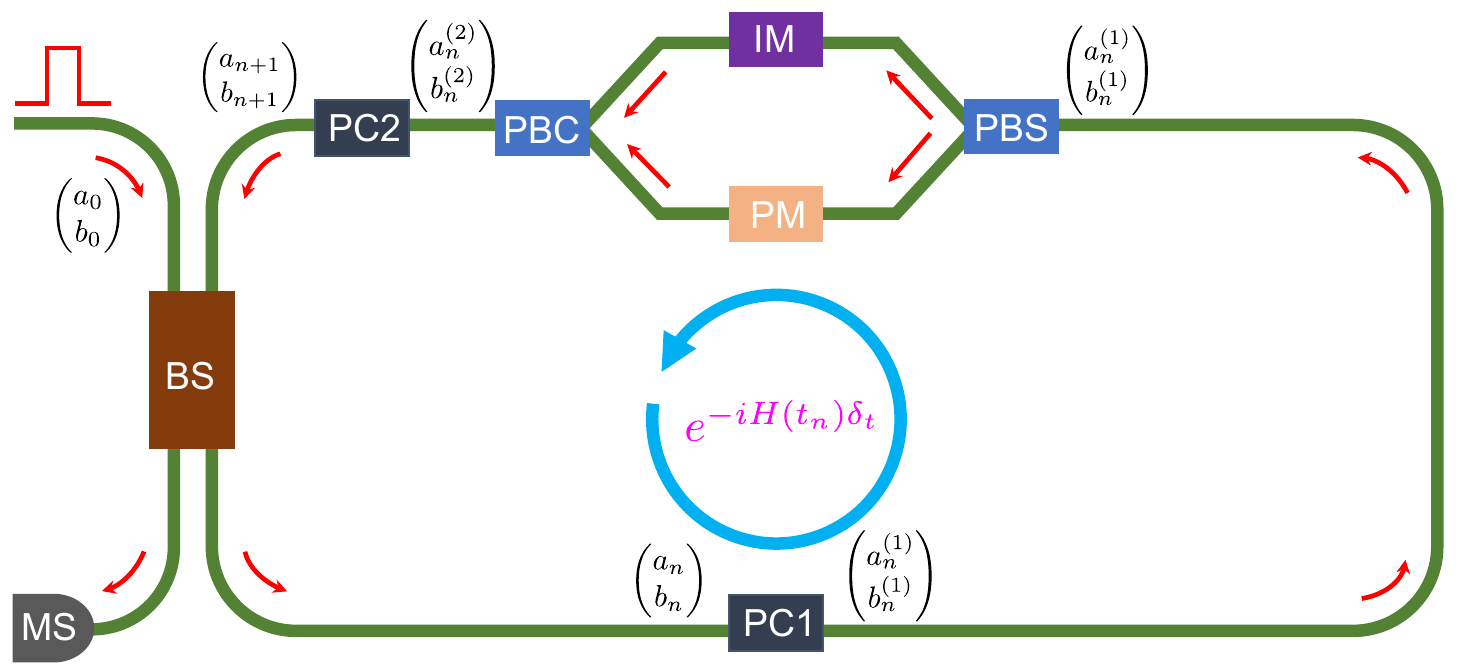}
    \caption{Schematic illustration of the fiber-loop arrangement for the full observation of adiabatic state evolution around the EP. BS: Beam splitter, PC: Polarization controller, PBS: Polarization beam splitter, IM: Intensity modulator, PM: Phase modulator, PBC: Polarization beam combiner and MS: Measurement system. A single loop realizes the time evolution operator $e^{-iH(t_n)\delta_t}$ for a small segment, and multiple round trips collectively implement the full evolution $U(t)$.}
    \label{figure3}
\end{figure}

Recently, increasingly refined time-control techniques in photonic simulators have been explored~\cite{NLL+22}, offering new insights into the characterization of the full dynamics of physical systems.
Then, we further refine our experimental setup by extending it into a fiber-loop structure, thereby additionally proposing an experimental scheme that allows to monitor the adiabatic time evolution of light in a continuous round-trip direct-detection system. By applying time-controlled intensity and phase modulations along with polarization coupling, each round trip in the loop effectively corresponds to a distinct point in the parameter space.
Extensive numerical simulations of this non-Hermitian setup allow us to provide a more comprehensive depiction of adiabatic state transitions in such platforms. Further details are provided below.

To realize the full adiabatic evolution, we can decompose the total time-evolution operator $U(t)$ into a product of small-segment unitary operations.
Each segment is approximated as $U_n=e^{-iH(t_n)\delta_t}$, where $n=1,2,\dots,N$,  and $\delta_t$ is the time interval of each segment.
We first perform a general decomposition of $U$ as:
\begin{eqnarray}\label{SVDn}
U_n = R_n^{1} L_n R_n^{2},
\end{eqnarray}
where $n=1,2,\dots,N$, $R_n^{1}$ and $R_n^{2}$ are $2\times2$ unitary operators, and $L_n$ is a $2\times2$ diagonal operator.
The following describes how to implement it: As illustrated in Supplementary Fig.~\ref{figure3}, the initial eigenstate can be encoded in the polarization state of the light pulse $|\Psi(0)\rangle=[a_0,b_0]^{T}$, and then coupled into the fiber loop via the beam splitter (BS). Next, the polarization controller (PC1) can realize $R_n^{1}$; this is because the polarization controller can be equivalently represented as a sandwiched combination of two quarter-wave plates and one half-wave plate (Q-H-Q), which can achieve the required $2\times2$ unitary transformation.

According to the Jones matrix form of the half-wave plate in Eq.~(9)
and the Jones matrix form of the quarter-wave plate in Eq.~(10), the optical pulse after undergoing the operation of PC1 can be universally expressed as
\begin{eqnarray}\label{SVDn}
\begin{pmatrix}
a_n^{(1)} \\
b_n^{(1)}
\end{pmatrix}
=
R_n^1(\varsigma_2,\varphi_1,\varsigma_1)
\begin{pmatrix}
a_n \\
b_n
\end{pmatrix},
\end{eqnarray}
where $R_n^{1}(\varsigma_2,\varphi_1,\varsigma_1)=U_q(\varsigma_2) U_h(\varphi_1) U_q(\varsigma_1)$.
Then we use a polarization beam splitter (PBS) to
guide the horizontal and vertical field components to two
separate branches, where they are independent but simultaneously undergo different gain or loss modulation, thus enabling us to achieve a diagonal $2\times2$ matrix $L_n$. These respective modulations are realized by introducing loop-dependent gain and phase modulations to these polarization components,
\begin{equation}
\begin{pmatrix}
a_n^{(2)} \\
b_n^{(2)}
\end{pmatrix}
=L_n\begin{pmatrix}
a_n^{(1)} \\
b_n^{(1)}
\end{pmatrix}=
\begin{pmatrix}
g_n & 0 \\
0 & p_n
\end{pmatrix}
\begin{pmatrix}
a_n^{(1)} \\
b_n^{(1)}
\end{pmatrix},
\label{eq:S2}
\end{equation}
Then, the two components are recombined using a polarization beam combiner (PBC). Afterwards, another polarization controller (PC2) realizes the unitary operator $R_n^{2}(\varsigma_4,\varphi_2,\varsigma_3)$ in a similar way. The full transformation in a single round trip is thus given by
\begin{equation}
\begin{pmatrix}
a_{n + 1} \\
b_{n + 1}
\end{pmatrix}
=
U_n \begin{pmatrix}
a_n \\
b_n
\end{pmatrix}=
R_n^{2}(\varsigma_4,\varphi_2,\varsigma_3)
\begin{pmatrix}
g_n & 0 \\
0 & p_n
\end{pmatrix}
R_n^1(\varsigma_2,\varphi_1,\varsigma_1)
\begin{pmatrix}
a_n \\
b_n
\end{pmatrix}.
\end{equation}
Ultimately, by sequentially modulating the optical pulse over
$N$ round trips in the fiber loop, we effectively realize the time-evolution operation $U(t)$.

\begin{figure}
    \includegraphics[width=0.95\linewidth]{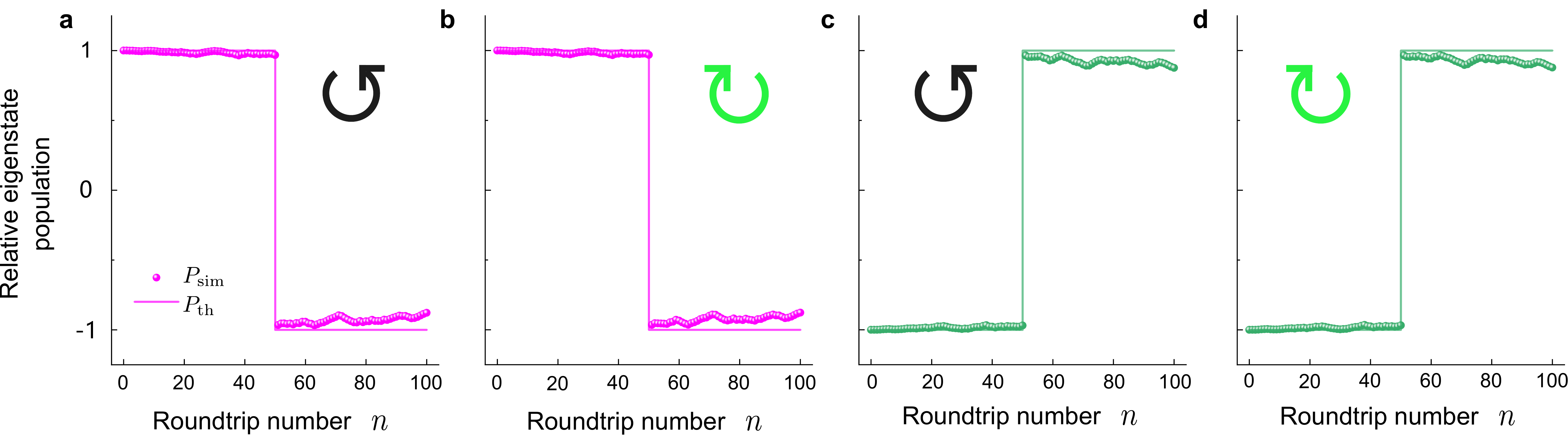}
    \caption{Numerical simulations of the proposed protocol for continuous observation of adiabatic state transfer. {\bf a-b} The initial state is chosen as $|\psi_1\rangle$. {\bf c-d} The initial state is chosen as $|\psi_2\rangle$. The relative eigenstate population $P_{\text{sim}}(t)$ is obtained based on our proposed scheme with a dynamic encirclement of the EP. Solid curves denote theoretical prediction results.
    We set the total roundtrip number $N=100$, meaning the full encircling period $T = 2\pi/\omega$ is discretized into 100 segments. Each roundtrip corresponds to a time step $\delta_t =T/N$.
    We note an abrupt change in the instantaneous eigenstates at $t=T/2$ in Fig.~2 of the main text, which precisely corresponds to the moment the evolution trajectory adiabatically transfers from one energy surface to another. Here, the same behavior is observed: the state transfer always occurs earliest at $t=T/2$, which corresponds to the roundtrip number $N/2=50$. The discrepancy between the simulation and theoretical results stems from the accumulation of errors.
    Other parameters are the same as those used in Fig.~2.
}
    \label{figureS1}
\end{figure}

We emphasize that the scheme presented here differs from our experimental setup in Fig.~1b of the main text. In Fig.~1b of the main text, each experiment implements a complete time-evolution operator $U(t)$, corresponding to one of $21$ discrete time points. The design of each $U(t)$ is constructed from the product of multiple segments, and only the final state is measured per run.
It also differs from standard spatially non-uniformly coupled waveguide systems, where information about the optical state is only available at the input and output ports.
In contrast, the proposed scheme enables a full simulation of the adiabatic evolution by directly implementing each small segment $U_n = e^{-iH(t_n)\delta_t}$ in a sequential manner.
In this scheme as illustrated in Supplementary Fig.~\ref{figure3}, at the end of each round trip, half of the pulses couple out of the optical fiber loop for monitoring purposes, while the remaining pulses remain in the loop to repeat the process until the required number of pulse cycles is reached. Each complete loop can realize the time evolution $e^{-iH(t_n)\delta_t}$ of a small segment, and multiple round trips can implement $U(t)$, i.e., each round trip in the fiber loop realizes a single segment, and the total evolution is reconstructed through repeated cycling. This cumulative approach allows us to faithfully capture the continuous evolution dynamics, including finer features of the adiabatic process.
However, this implementation requires significantly greater experimental complexity. Precise temporal control is essential for both phase and amplitude modulators triggered by a time-related arbitrary waveform generator, as well as polarization management. As a result, high-speed, high-precision instrumentation is necessary to ensure the fidelity of the evolution.

We additionally conduct numerical simulations of the proposed single-fiber experimental protocol. For that, we perform $100$ iterations of the operation with roundtrip number $N$ set to $100$. As shown in Supplementary Fig.~\ref{figureS1}, the simulation results clearly demonstrate the adiabatic state transfer, and we observe good agreement with the theoretical results, which significantly improves the simulation of adiabatic state evolution.
This behavior can be well understood: a higher roundtrip count corresponds to a smaller time interval, thereby enhancing simulation accuracy and better approximating the slow time-dependent adiabatic evolution process.

As shown in the Supplementary Fig.~\ref{figureS2}, based on the fiber loop simulator we previously discussed, we also numerically simulate the state transfer dynamics for a static initial eigenstate, and find that mutual transfer between the two initial states also occurs, regardless of the direction of encirclement around the EP.
\begin{figure}[h]
    \includegraphics[width=0.95\textwidth]{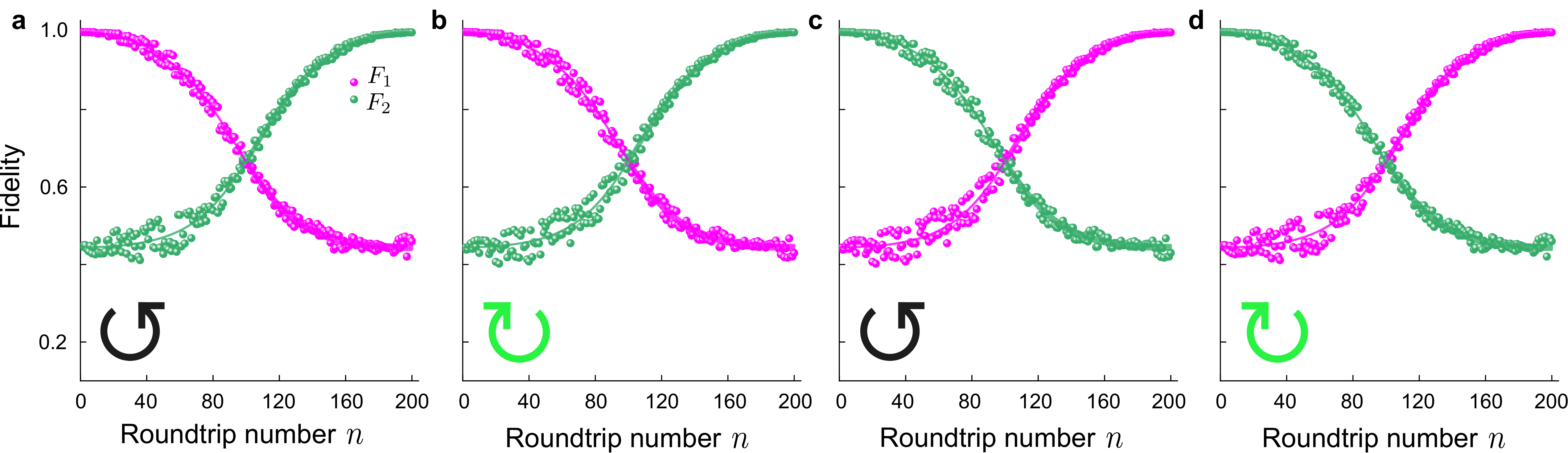}
\caption{ Numerical simulations of the adiabatic state transfer related to the initial right eigenvector while dynamically encircling around an EP.
We set the total roundtrip number $N=200$. {\bf a-b} The initial state is chosen as $|\psi_1\rangle$. {\bf c-d} The initial state is chosen as $|\psi_2\rangle$.
$F_{k}=|\langle \psi_k(t=0)|\Psi_\text{sim}(t)\rangle|^2$ ($k=1,2$) is the fidelity between the simulated time-evolving right eigenstate $|\Psi_\text{sim}(t)\rangle$ based on the scheme and static right eigenvector, i.e., the initial state of the system $|\psi_k(t=0)\rangle$. Solid curves are obtained from fidelity between the theoretical time-evolving right eigenstate $|\Psi_\text{th}(t)\rangle$ and the initial state of the system.}
    \label{figureS2}
\end{figure}

\subsection{Supplementary Note 8. Evolution of the Polarization Components in the Measured States}

In this section, we plot the evolution of the Stokes vector components ($S_1, S_2, S_3$)  of the experimentally measured states, providing a complete characterization of the state transfer evolution under different encirclements around the EP.

\begin{figure}[t]
    \includegraphics[width=0.99\textwidth]{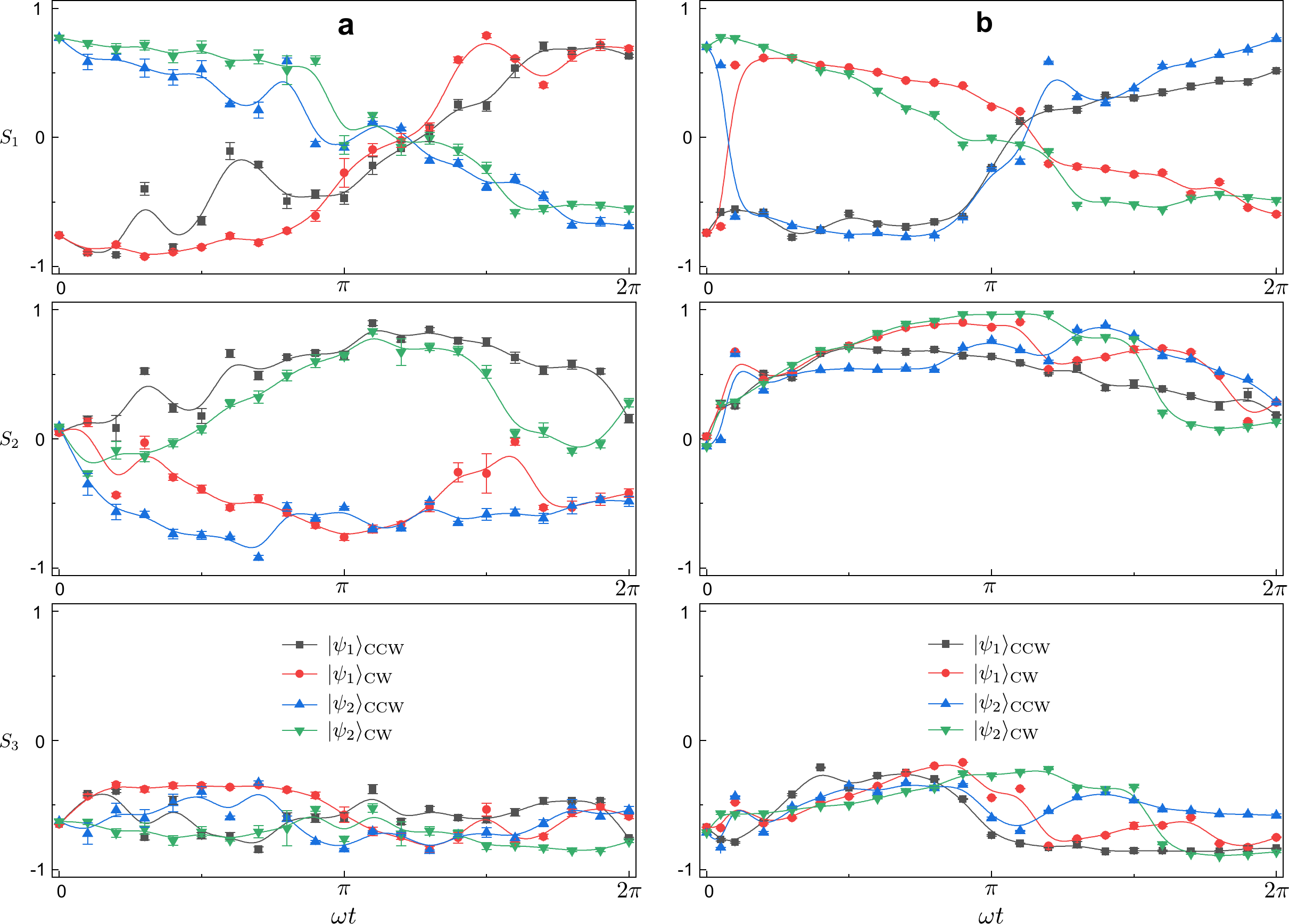}
\caption{Evolution of the Stokes vector components components $S_1, S_2,$ and $S_3$ during ($\mathbf{a}$) adiabatic state transfer and ($\mathbf{b}$) chiral state transfer. Data points represent experimental measurements for two distinct initial states, $|\psi_1\rangle$ (black and red) and $|\psi_2\rangle$ (blue and green), and the solid curves are obtained by smoothly fitting the experimental data. The markers labeled ``CW'' and ``CCW'' indicate clockwise and counter-clockwise encirclements around the EP, respectively.
The trajectories of evolution illustrate the state transfer process within the adiabatic/non-adiabatic framework, where the final state distributions reflect the geometric properties of the system. The deviations are dominated by experimental imperfections in implementing the non-unitary evolution operator and measurement uncertainties. Other parameters are the same as those used in Fig.~2.}
    \label{figureS_stokes}
\end{figure}

Supplementary Figure~\ref{figureS_stokes}a illustrates the adiabatic state transfer. In this regime, the evolution is largely symmetric with respect to the encirclement direction, demonstrating the robustness of the adiabatic mapping. The $S_1$ component, in particular, demonstrates a robust adiabatic-symmetric transfer regardless of the encircling direction, which is closely mirrored by the dynamics observed in Fig.~2 and Supplementary Fig.~\ref{figureS_fid}.
In contrast, Supplementary Fig.~\ref{figureS_stokes}b depicts the chiral state transfer dynamics. Here, we observe a distinct breaking of the winding symmetry in the state transfer. Specifically, the final configuration of the component $S_1$ depends fundamentally on the chirality (direction) of the encirclement around the EP, which is independent of the initial state. This behavior is in good agreement with the results presented in Fig.~3.

\subsection{Supplementary Note 9. Perturbation Analysis of the Observed Adiabatic Mode Switching}
\label{sec:stability_analysis}

In this section, we analyze the impact of perturbations of the NHH on adiabatic state transfer during the dynamic encirclement of an EP. The system's state evolution can change drastically depending on the nature of the Hamiltonian perturbations, primarily because such perturbations can render the NHH eigenvalues complex-valued. This may induce NATs in the system dynamics.

We note that a simple uniform imaginary shift of the NHH eigenvalues, i.e., $H''(t) = H(t) + i\Delta I$, where $I$ is the identity matrix and $\Delta \in \mathbb{R}$, does not affect adiabatic state transfer. For a NAT to occur, the imaginary parts of the two eigenvalues must, in general, be different. This difference ensures that during the dynamic EP encirclement, the system may undergo a non-adiabatic transition, favoring the eigenstate belonging to the energy manifold with the larger imaginary part.

Consider instead perturbing the NHH in Eq.~(3) as
\begin{equation}
    H''(t) = H(t) + \begin{pmatrix}
        \Delta & 0 \\
        0 & 0
    \end{pmatrix},
    \label{eq:perturbed_H}
\end{equation}
which yields the eigenenergies $E''_{1,2} = \frac{1}{2}\left( \Delta \mp \sqrt{\Delta^2 + 4\alpha^2 + 4\alpha\Delta\cos\theta} \right)$.
Thus, the difference between the imaginary parts of the eigenvalues, $\operatorname{Im}(E''_2-E''_1)$, increases with $|\Delta|$. This increase can eventually lead to the emergence of NATs and, specifically, to chiral mode behavior.

We illustrate the effect of this perturbation on the state-transfer fidelity in Supplementary Figs.~\ref{figureS_scan_per_1234}-\ref{figureS_scan_per_5678} for different values of $\Delta$. The relative eigenstate population is obtained identically to that shown in Fig.~2 of the main text.

\begin{figure}
    \includegraphics[width=0.95\textwidth]{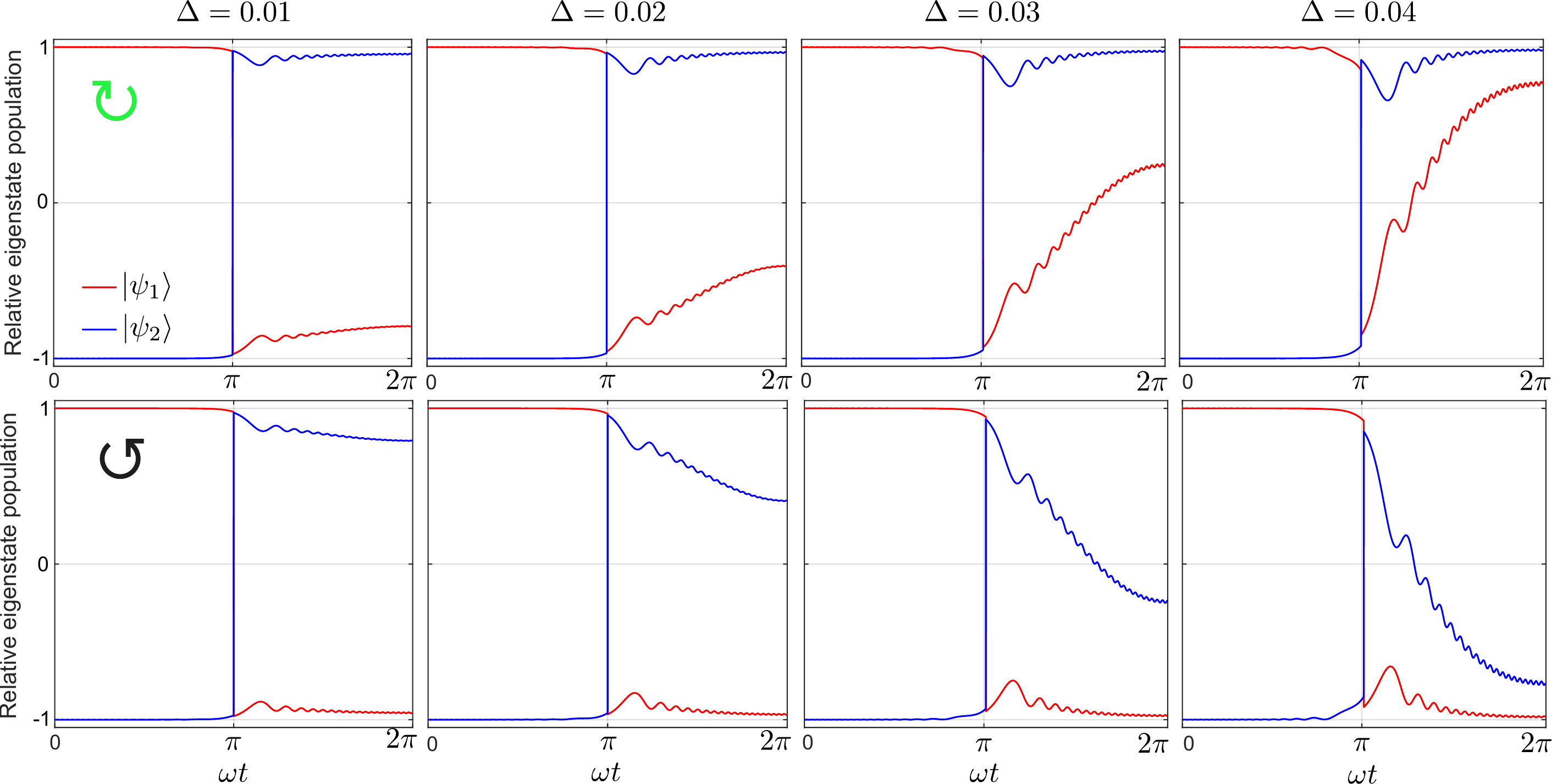}
\caption{Relative eigenstate population $P(t)$ for clockwise (top row) and counter-clockwise (bottom row) parameter encirclements for different $\Delta$ while encircling the EP for Eq.~\eqref{eq:perturbed_H} at $\omega=\pi/100$. The red curves represent the evolution starting from the initial state $|\psi_1\rangle$, while the blue curves denote the evolution from $|\psi_2\rangle$. Other descriptions are the same
as those used in Fig.~2.}
    \label{figureS_scan_per_1234}
\end{figure}

\begin{figure}
    \includegraphics[width=0.95\textwidth]{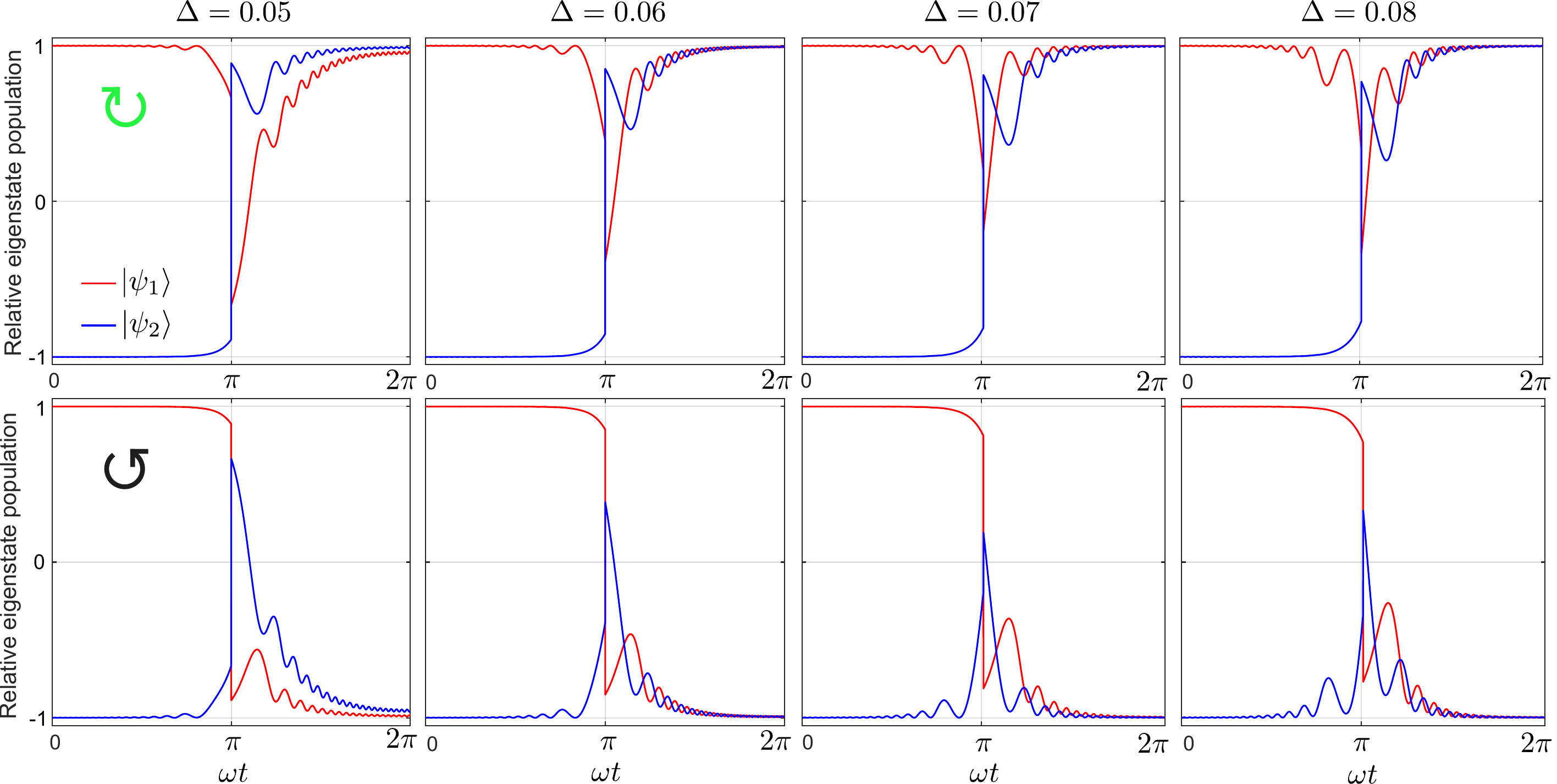}
\caption{Relative eigenstate population $P(t)$ for clockwise (top row) and counter-clockwise (bottom row) parameter encirclements for different $\Delta$ while encircling the EP for Eq.~\eqref{eq:perturbed_H}. Other descriptions are the same
as those used in Supplementary Fig.~\ref{figureS_scan_per_1234}.}
    \label{figureS_scan_per_5678}
\end{figure}

\begin{figure}[h!]
    \includegraphics[width=0.80\textwidth]{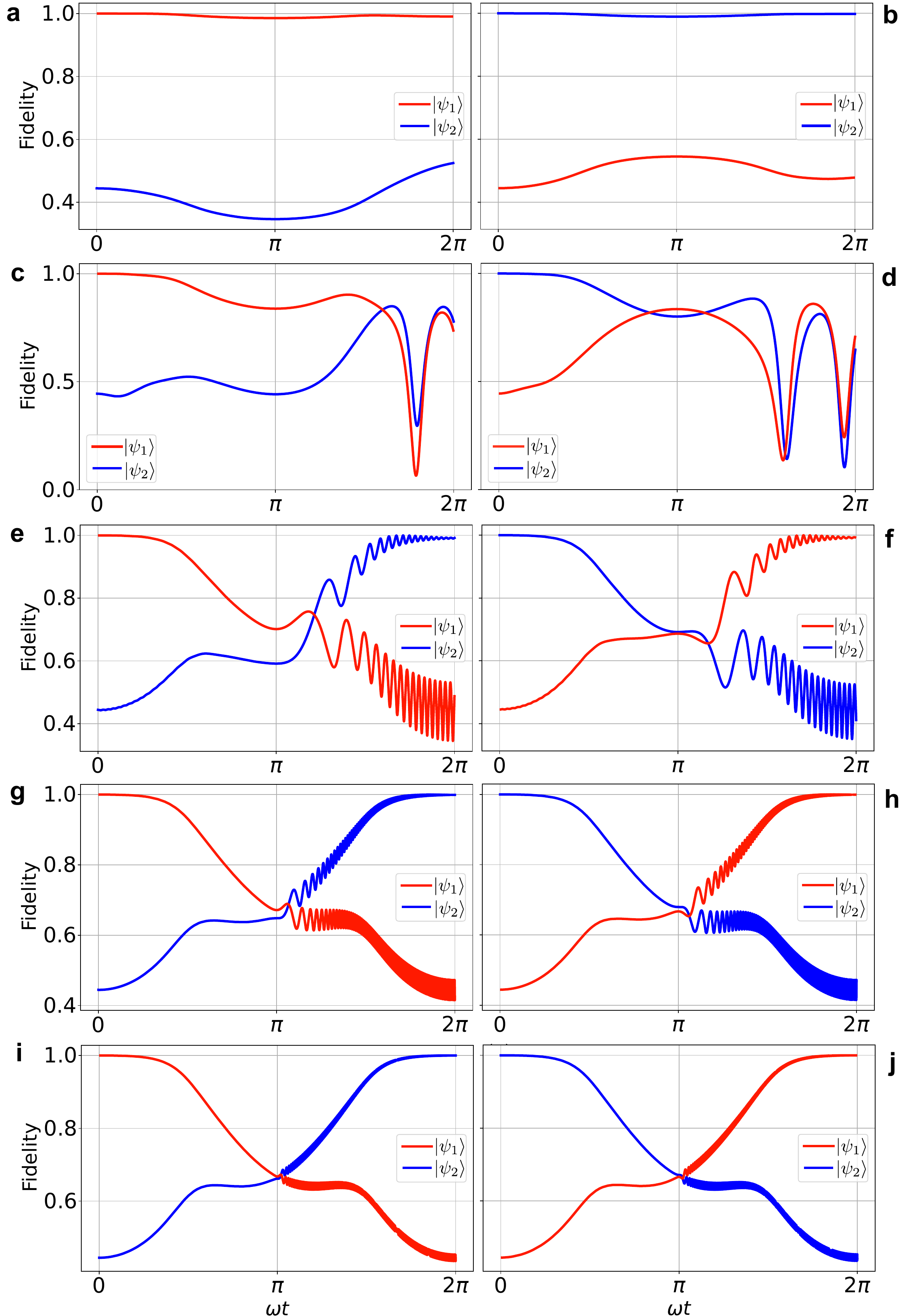}
\caption{Similar to Fig.~S3, but obtained from numerical simulations, showing the fidelity between the evolving and initial states for different values of the winding period $T$, irrespective of the winding direction. The left column corresponds to the case in which the initial state is chosen as $|\psi_1\rangle$, while the right column corresponds to $|\psi_2\rangle$. Each row corresponds to the EP-encircling protocol for the given period $T$, Namely, {\bf a-b} $T=1$ [arb. units], {\bf c-d} $T=10$ [arb. units], {\bf e-f} $T=100$ [arb. units], {\bf g-h} $T=1000$ [arb. units], {\bf i-j} $T=10000$ [arb. units]. The chosen loop trajectory around the EP is defined by the equations: $x=0.2\sin(\omega t+\pi)$, and $y=1-0.5\cos(\omega t+\pi)$. For this specific figure, the numerical simulations were performed using the Runge–Kutta method. The rapid oscillations observed in the second half of the winding period originate from the numerical stiffness encountered when crossing the diabolic line (further technical details regarding this solution stiffness can be found in Ref.~\cite{arkhipov2025}, see Appendix~G therein).
}
    \label{figureS_scan_T}
\end{figure}

\begin{figure}[h!]
    \includegraphics[width=0.95\textwidth]{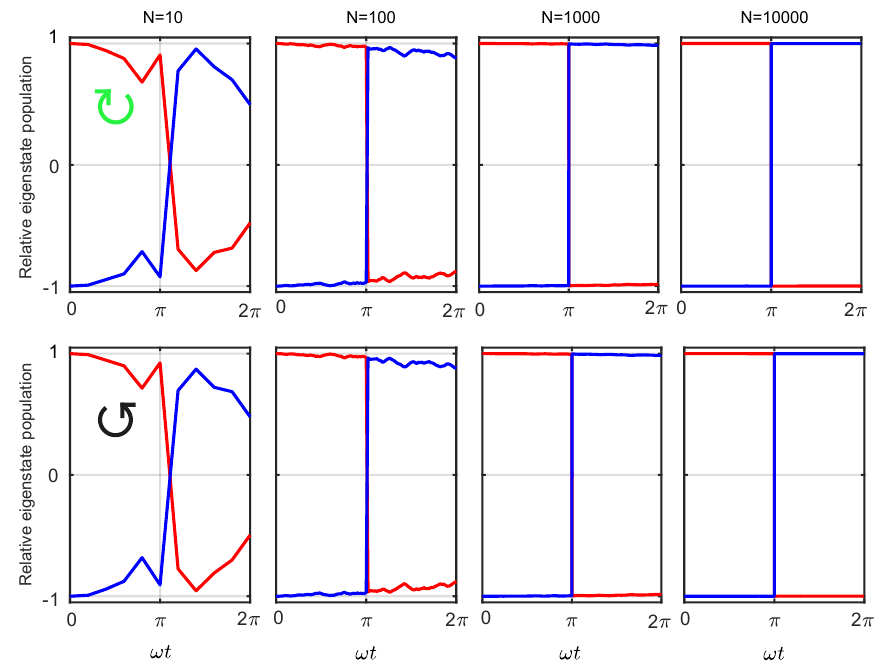}
\caption{Convergence and robustness analysis of the adiabatic protocol for the discretization steps $N$. Numerical simulations of the relative eigenstate population $P(t)$ as a function of the discretization steps $N$ and evolution speed $\omega$ for different $N$ values ranging from $10$ to $10^4$ at a fixed speed $\omega = \pi/50000$. The red curves represent the evolution starting from the initial state $|\psi_1\rangle$, while the blue curves show the evolution from $|\psi_2\rangle$. The top row displays the results for clockwise (CW) encirclement, and the bottom row shows the results for counter-clockwise (CCW) encirclement. The results show that $P(t)$ converges rapidly to unity as $N$ increases, with fluctuations becoming negligible for $N \ge 1000$. Other parameters are the same as those used in Fig.~2.}
    \label{figureS_scan_N}
\end{figure}

\begin{figure}[h!]
    \includegraphics[width=0.95\textwidth]{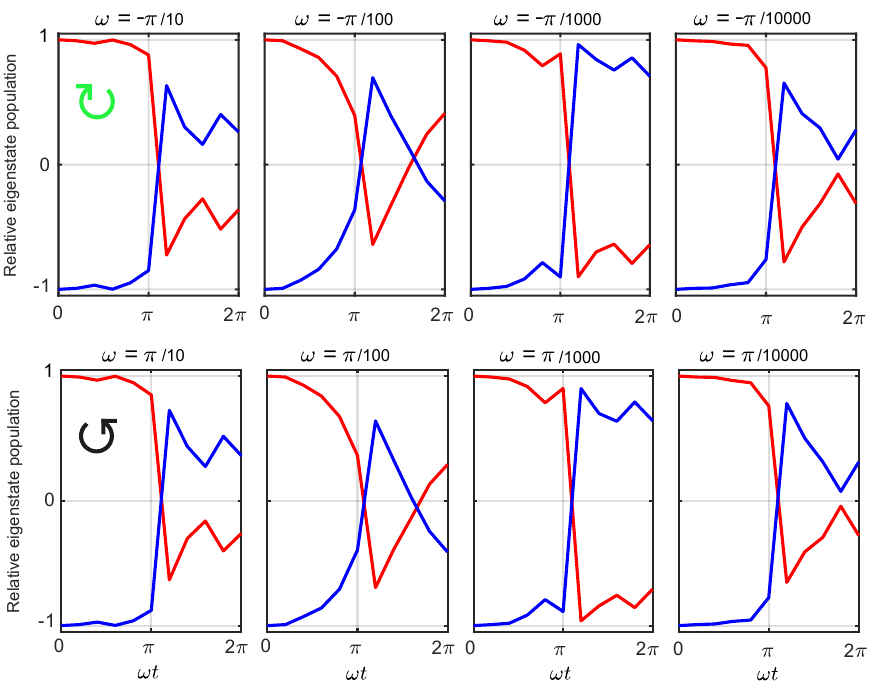}
\caption{Numerical simulations of the relative eigenstate population $P(t)$ for various evolution speed $\omega$ and $N=10$.}
    \label{figureS_scan_w_N=10}
\end{figure}

\begin{figure}[h!]
    \includegraphics[width=0.95\textwidth]{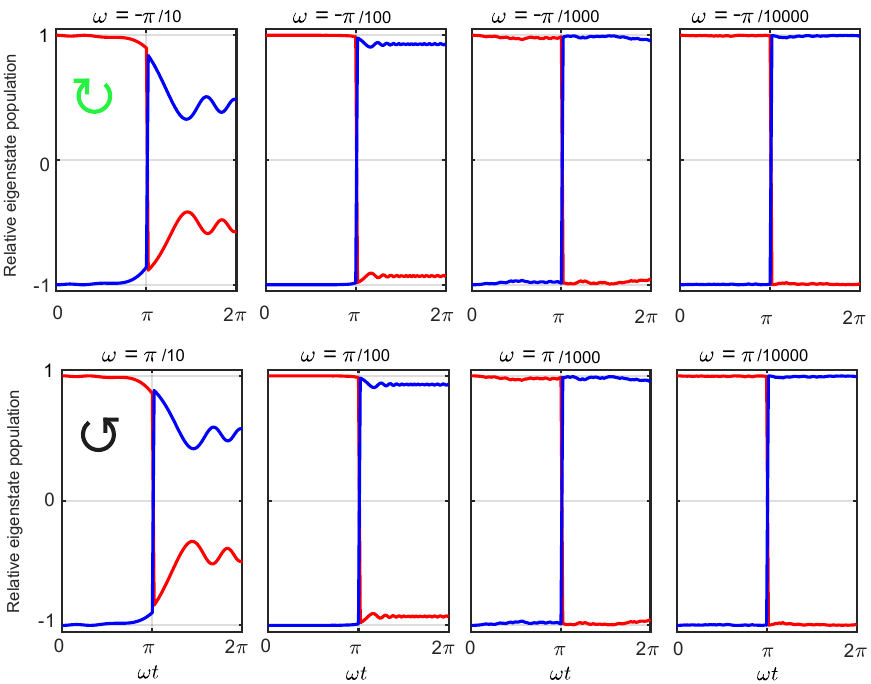}
\caption{Numerical simulations of the relative eigenstate population $P(t)$ for various evolution speed $\omega$ and $N=100$.}
    \label{figureS_scan_w_N=100}
\end{figure}

\begin{figure}[h!]
    \includegraphics[width=0.95\textwidth]{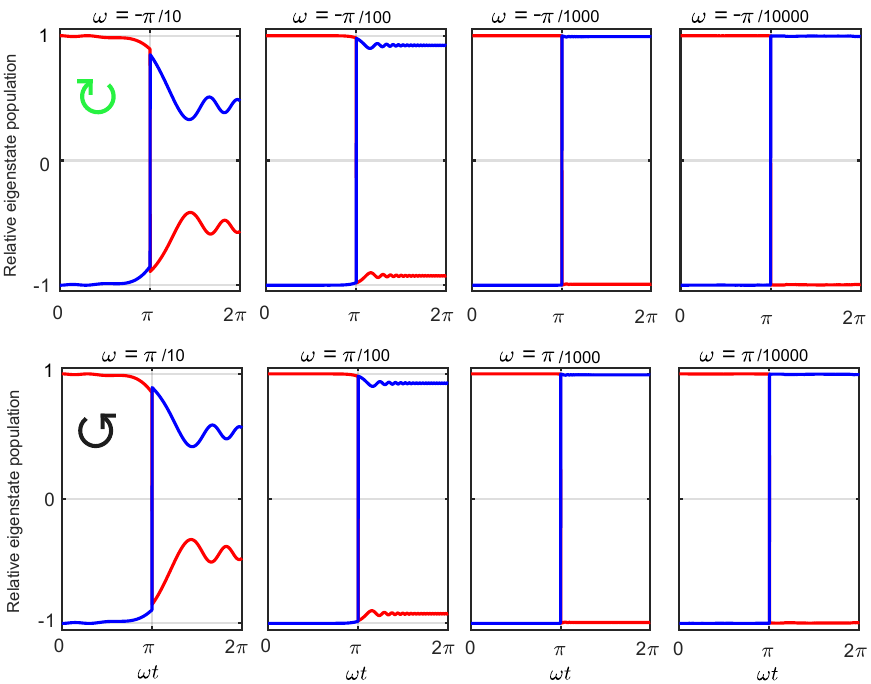}
\caption{Numerical simulations of the relative eigenstate population $P(t)$ for various evolution speed $\omega$ and $N=1000$.}
    \label{figureS_scan_w_N=1000}
\end{figure}

\begin{figure}[t!]
    \includegraphics[width=0.95\textwidth]{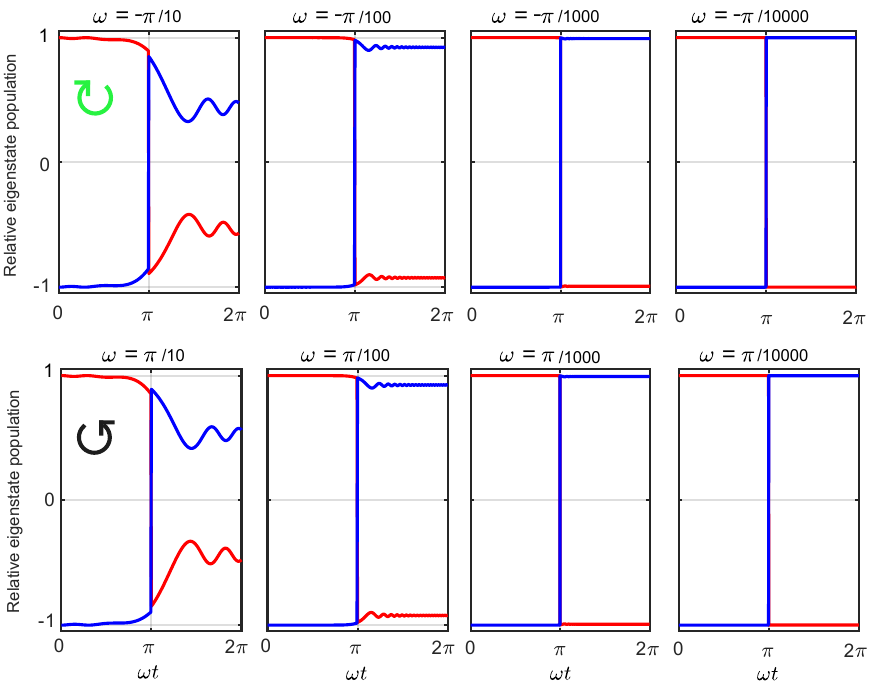}
\caption{Numerical simulations of the relative eigenstate population $P(t)$ for various evolution speed $\omega$ and $N=50000$.}
    \label{figureS_scan_w_N=50000}
\end{figure}

\begin{figure}[h!]
    \includegraphics[width=0.95\textwidth]{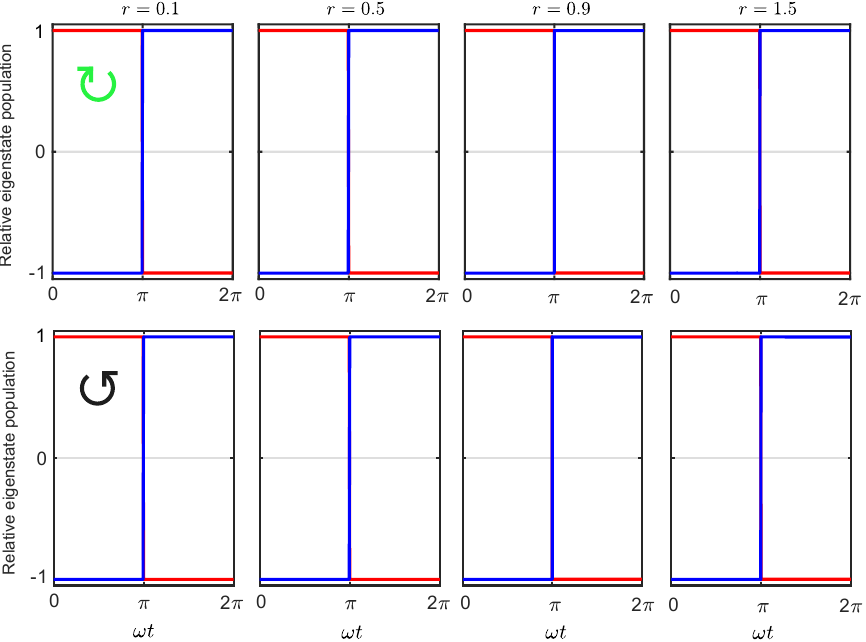}
\caption{Numerical simulations of the relative eigenstate population $P(t)$ for various encirclement radius $r$ around the EP, with $N=50000$ and $w=\pi/50000$. The top row displays the results for CW, and the bottom row for CCW.}
    \label{figureS_scan_r}
\end{figure}

\begin{figure}[h!]
    \includegraphics[width=0.95\textwidth]{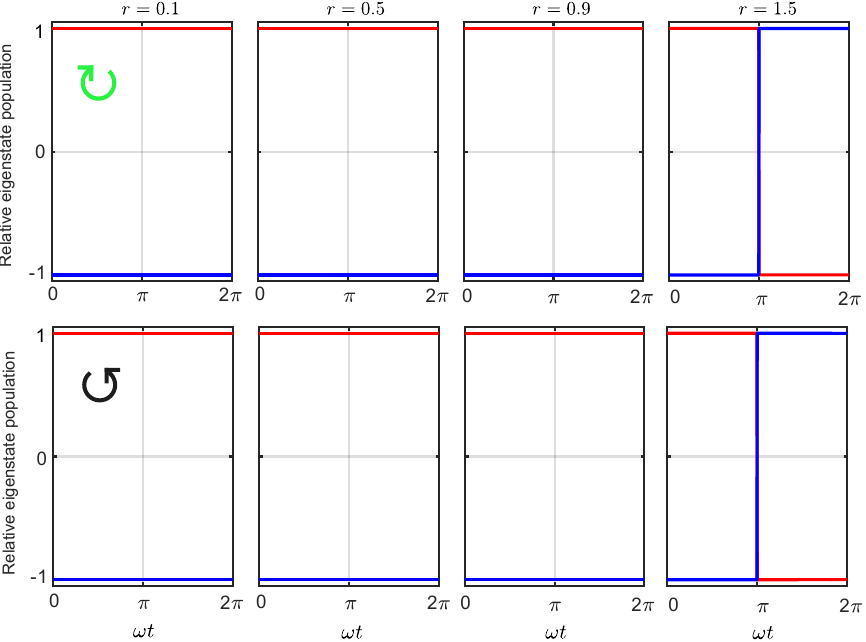}
\caption{Numerical simulations of the relative eigenstate population $P(t)$ for the encirclement center is at (0,2), with $N=50000$ and $w=\pi/50000$. The top row displays the results for CW, and the bottom row for CCW.}
    \label{figureS_scan_center=2}
\end{figure}

As shown in Supplementary Fig.~\ref{figureS_scan_per_1234}, adiabatic state transfer persists for very small ($\Delta \leq 0.03$). However, the evolving states begin to acquire an exponential factor due to the loss of pseudo-Hermiticity in the perturbed NHH $H''$, whose eigenvalues become complex. This exponential factor results either in damping or amplification of the evolving state.
However, at larger perturbation strengths ($\delta \geq 0.05$), adiabaticity breaks down, leading to NATs and consequently to chiral mode behavior, as demonstrated in Supplementary Fig.~\ref{figureS_scan_per_5678}.

\subsection{Supplementary Note 10. Impact of Evolution Speed and Loop Parameters on Restored Adiabaticity and Robustness Analysis}

In this section, we present additional numerical results supporting our experimental observations of adiabatic state dynamics. By analyzing the interplay between the discretization step $N$, the evolution speed $\omega$, and the loop parameters, we numerically show that the observed adiabatic state transfer is robust and reliably reproduced across a wide range of conditions.

Firstly, we plot the new numerically simulated fidelity in Supplementary Fig.~\ref{figureS_scan_T}, which complement the experimental data shown in Fig.~S3. These simulations present the fidelity of the evolving state for different values of the winding period $T$. The newly added plots demonstrate that, in the adiabatic (slow-driving) limit $T\to\infty$, the state evolution remains continuous and the state-switching protocol is robust. By contrast, for sufficiently short winding periods, diabatic effects emerge, leading to deviations from the ideal evolution along the eigenenergy manifolds and a breakdown of the state-switching protocol.

Next, we examine the role of the discretization step $N$ used to obtain the experimentally implemented non-unitary operator $U(t)$ in the simulated dynamics. Setting the evolution speed to $\omega = \pi/50000$, we computed the relative eigenstate population $P(t)$ as a function of $N$, ranging from $10$ to $10^4$, as shown in Supplementary Fig.~\ref{figureS_scan_N}. The results show that $P(t)$ rapidly converges to unity as $N$ increases, with fluctuations becoming negligible for $N \gtrsim 1000$. For the experimental choice of $N = 50,000$ (used in the main text), the population is effectively constant, reflecting highly stable adiabatic state transfer. The consistent behavior observed in both the top and bottom rows further highlights the inherent symmetry of the protocol. These findings confirm that, unlike chiral EP encirclement, the restored adiabaticity is optimized in the adiabatic limit ($\omega \to 0$) and remains robust with respect to the encirclement direction. In contrast, for small $N$, the adiabatic behavior deteriorates significantly and may even vanish entirely.

To further explore the role of the evolution speed $\omega$ (with total evolution time $T = 2\pi/\omega$) on the adiabatic behavior, we numerically calculate $P(t)$ for various $\omega$ at fixed values of $N$, as shown in Supplementary Figs.~\ref{figureS_scan_w_N=10}–\ref{figureS_scan_w_N=50000}. The results demonstrate that the evolution speed critically influences adiabaticity: slower evolution (smaller $\omega$) promotes more pronounced adiabatic behavior, whereas even with large $N$, adiabaticity is suppressed if $\omega$ is too large. As $N$ increases ($N>100$) and $\omega$ decreases ($\omega<\pi/100$), the system rapidly converges to a highly stable and symmetric adiabatic transfer, independent of the winding direction, confirming the robustness of the protocol in the high-$N$ and low-$\omega$ limits.

As shown in Supplementary Figs.~\ref{figureS_scan_r}-\ref{figureS_scan_center=2}, we also numerically evaluate the robustness of the scheme by varying the loop parameters, which control the ``size'' (the radius $r$ and center) of the trajectory in parameter space.
In Supplementary Fig.~\ref{figureS_scan_r}, the center of the trajectory is located at the EP, $(x,y)_{\text{EP}} = (0,1)$. One can see that, regardless of the value of the radius $r$, the state evolution exhibits adiabatic characteristics. While the center of the trajectory is located at $(x,y)= (0,2)$ in Supplementary Fig.~\ref{figureS_scan_center=2}, only when $r>1$ and the trajectory precisely encircles the EP, does adiabatic state exchange occur.
These results demonstrate that as long as the trajectory remains on the real-spectrum hyperboloid manifold and encircles the EP, the symmetric adiabatic transfer is preserved. This highlights the topological protection inherent in our scheme.

We thus confirm that the adiabatic state transfer is observed for both clockwise and counter-clockwise encirclement directions, highlighting the intrinsic symmetry of the protocol. Taken together, these numerical results provide clear support for the conclusion that the experimentally observed ``restored adiabaticity'' reflects a robust and reproducible physical phenomenon.


\end{document}